\documentclass[aps,prd,
reprint,
floatfix,
superscriptaddress,
nofootinbib,
showpacs
]{revtex4-1}

\usepackage{graphicx}
\usepackage{epstopdf}
\usepackage{amssymb}
\usepackage{amsmath}
\usepackage{setspace}
\usepackage{color}
\usepackage{comment}
\usepackage{hyperref}

\newcommand{\unit}[1]{\ensuremath{\ \mathrm{#1}}}
\newcommand{\unitns}[1]{\ensuremath{\mathrm{#1}}}
\newcommand{\dd}{\ensuremath{\mathrm{d}}}
\renewcommand{\vec}[1]{\mathbf{#1}}

\begin{document}

\title{Testing long-distance modifications of gravity to 100~astronomical units}

\author{Brandon Buscaino}
\affiliation{Department of Physics, Stanford University, Stanford, CA 94305, USA}
\author{Daniel DeBra}
\affiliation{Department of Aeronautics and Astronautics and HEPL, Stanford University, Stanford, CA 94305, USA}
\author{Peter W.~Graham}
\affiliation{Stanford Institute for Theoretical Physics, Department of Physics, Stanford University, Stanford, CA 94305, USA}
\author{Giorgio Gratta}
\affiliation{Department of Physics and HEPL, Stanford University, Stanford, CA 94305, USA}
\author{Timothy D.~Wiser}
\email{tdwiser@stanford.edu}
\affiliation{Stanford Institute for Theoretical Physics, Department of Physics, Stanford University, Stanford, CA 94305, USA}

\date{\today}

\begin{abstract}

There are very few direct experimental tests of the inverse square law of gravity at distances comparable to the scale of the Solar System and beyond.	Here we describe a possible space mission optimized to
test the inverse square law at a scale of up to 100~AU.	For example, sensitivity to a Yukawa correction with a strength of $10^{-7}$ times gravity and length scale of 100~AU is within reach, improving the current state of the art by over two orders of magnitude. This experiment would extend our understanding of gravity to the largest scale that can be reached with a direct probe using known technology.
This would provide a powerful test of long-distance modifications of gravity including many theories motivated by dark matter or dark energy.

\end{abstract}


\maketitle

\tableofcontents

\section{Introduction}

In recent times a great deal of attention has been devoted to searching for modifications of Newtonian
gravity at or below
mm distances~\cite{Adelberger:2003zx}.	This possibility is motivated for example by theories with large
extra dimensions or supersymmetric theories with light moduli~\cite{Dimopoulos:1996kp, ArkaniHamed:1998nn,
Antoniadis:1998ig, ArkaniHamed:1998rs}.
At the opposite extreme of the distance spectrum, deviations from Newtonian gravity at very long
distance scales may be related to astrophysical and cosmological problems in modern physics~\cite{Decadal_survey}.
The cosmological constant problem is one of the deepest and most enduring mysteries in modern physics. There have been many theories which attempt to modify
gravity on long distance scales, for example Dvali-Gabadadze-Porrati (DGP) gravity or theories of massive gravity
~\cite{Dvali:2000hr, Dvali:2002pe, Dvali:2002fz, ArkaniHamed:2002fu, deRham:2010ik, deRham:2010tw,
deRham:2010kj, D'Amico:2011jj, deRham:2012az}.
In these theories large effects are predicted to
arise only on cosmological length scales.  However, smaller effects may be seen at shorter, experimentally
accessible length scales \cite{Dvali:2002vf, Berezhiani:2013dw}.  Further, theories that replace dark matter
with modifications of Newtonian gravity, such as modified Newtonian dynamics (MOND)~\cite{MOND,MONDReview} may also lead to observable effects on long,
yet still accessible scales \cite{Blanchet:2011pv}.  The enduring negative results~\cite{CDMS,XENON,LUX} in the search of
dark matter in the form of weakly interacting massive particles (WIMPs) may indeed suggest a broader approach
to this important problem. Although theories with long-distance modifications of gravity have historically been troubled by ghosts, discontinuities, and other theoretical difficulties, recent progress has been made in alleviating these concerns; for a review, see Ref.~\cite{Hinterbichler:2011tt}. For these reasons, we focus on the directly observable consequences of long-distance modifications of gravity and estimate the potential of a dedicated space mission to measure modifications of the gravitational force of the Sun out to 100~AU.	This is the largest distance accessible for direct measurements in a practical amount
of time using known technology and would provide a powerful test of long-distance modifications of gravity.

\begin{figure}[t!]
\includegraphics[width=3.4in]{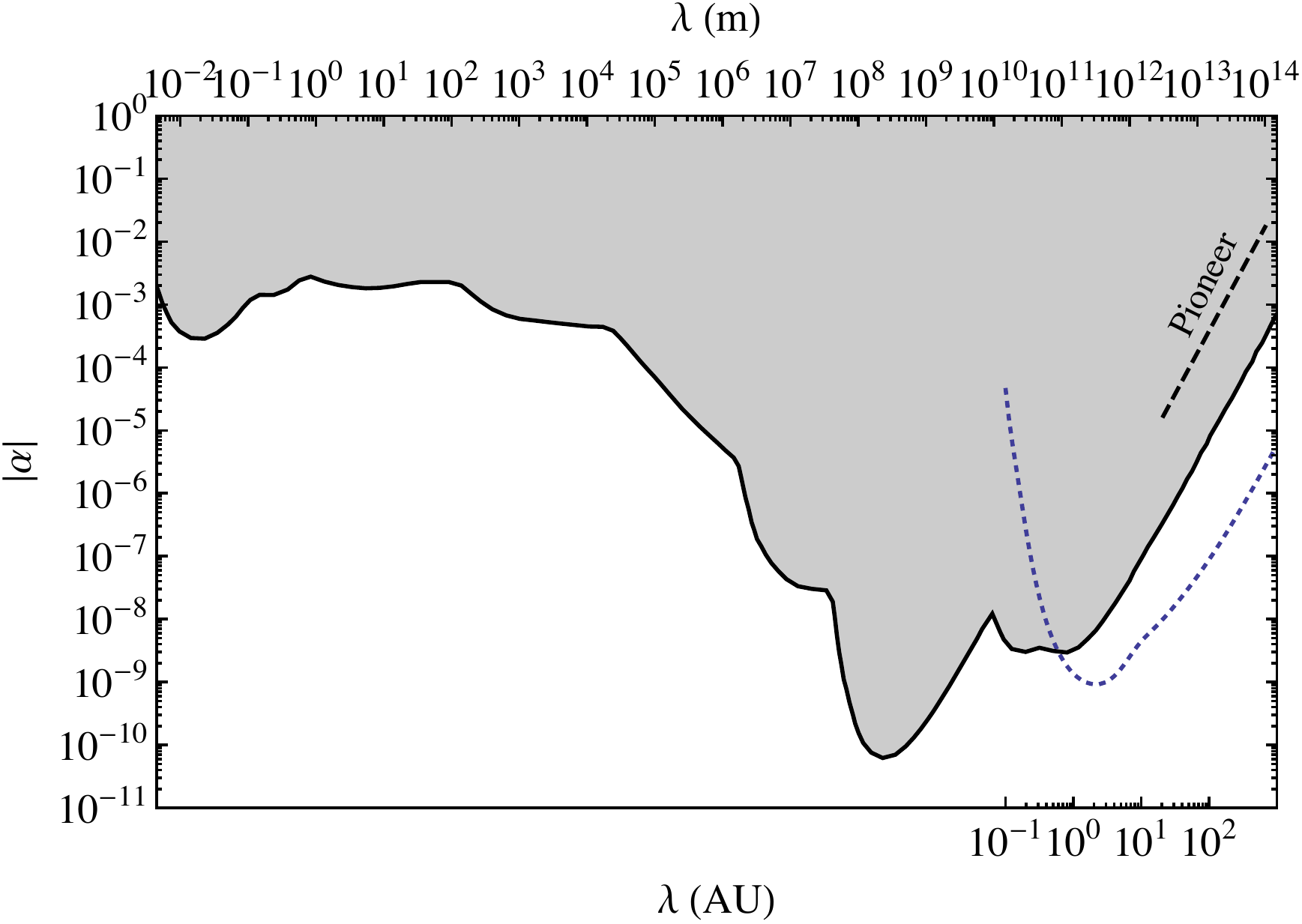}
\caption{Existing 2$\sigma$ experimental limits on new Yukawa forces with strength $| \alpha |$ relative to gravity as function of the scale $\lambda$.
The grey region, adapted from~\cite{Adelberger:2003zx}, is the current state of the art. The dashed line is the size of the ``Pioneer anomaly,'' which can be interpreted as a limit set by the trajectory of the Pioneer spacecraft. The dotted curve corresponds to the expected sensitivity of the experiment proposed here.}
\label{fig:alpha_lambda}
\end{figure}

One way to parametrize possible deviations from the $1/R^2$ behavior of gravity is by introducing a new Yukawa
force with charge proportional to mass, so that the effective gravitational potential can be written as
\begin{equation}
\Psi(R)=-\frac{GM}{R} \left[1+\alpha e^{-R/{\lambda}}\right],
\label{eq:Yuk}
\end{equation}
where $G$ is Newton's constant, $M$ is the source mass, and $R$ is the distance from the source. The new Yukawa interaction then has a strength $\alpha$ relative to gravity and a characteristic length scale $\lambda$.  In this framework, experiments measure (or constrain) the dimensionless parameter $\alpha$ as a function of
the distance scale $\lambda$.	A summary of current limits on the magnitude of $\alpha$ is
shown in Fig.~\ref{fig:alpha_lambda}.

Qualitatively, the sensitivity of experimental measurements to new Yukawa forces of strength $\alpha$ improves as $\lambda$
becomes substantially larger than Earth-scale inhomogeneities and reaches a level $< 10^{-10}$
for $\lambda \sim R_{\rm Moon}$ (the radius of the lunar orbit) owing to laser
lunar ranging measurements~\cite{Decadal_survey}.    Beyond such distance scales, the orbital mechanics of
the planets in the Solar System forms the most stringent test of new Yukawa contributions to gravity~\cite{talmadge}.
The Pioneer 10 and 11 spacecraft provided an alternative way to measure $\alpha$ at very long scales $\lambda\sim 10$--100 AU as they
receded from the Sun over a period of over 30~years.   Indeed for some time the analysis
of the Pioneer data appeared to indicate an anomalous acceleration towards the Sun of
$10^{-10}$~m/s$^2$~\cite{PioneerAnomaly,Anderson:2001sg}, resulting in $|\alpha| \sim 1.7\times 10^{-4}$ at
$\lambda \sim 10^{13} $~m.  A subsequent, more careful analysis, however, attributed this effect to instrumental
systematics~\cite{Pioneer_resolved}. The many uncertainties resulting from spacecraft
designed for other purposes make the upper limit on $| \alpha |$ not competitive with that
computed from planetary dynamics.

As will be discussed (and as shown in Fig.~\ref{fig:alpha_lambda}), a specially optimized space mission offers the opportunity of improving sensitivity to a new Solar System-scale Yukawa force by at least two orders of magnitude
with respect to the present state of the art. We show that such a mission can be relatively
simple and would use only well-understood and tested technology.  In addition, measurements directly
done along a trajectory actually reaching 100~AU constitute a test that is fully model independent and
would detect anomalies that are not well-described by a Yukawa term.  This is in contrast with
measurements using planetary motion that rely on the specific Yukawa parametrization to extract
deviations at scales $\lambda$ different from the orbital radius.
While the Yukawa potential is quite generic, there are interesting modifications of gravity that produce truly long-distance effects that are not accurately described by a Yukawa potential. Reliance on the Yukawa functional form is a limitation of the previous searches in Ref.~\cite{talmadge}. In a strict sense, no data exists beyond the distance of Jupiter except for the limits set by the Pioneer spacecraft.  The mission proposed here would improve these direct limits by over four orders of magnitude by carefully controlling systematic effects and performing direct measurements along the journey to 100~AU.

\section{Experimental concept}

Two guiding principles are key in the design of a space mission optimized for a sensitive
search for deviations from the $1/R^2$ law at large distances: the mission should have a
reasonable duration and the spacecraft should be designed in such a way as to minimize the
non-gravitational interactions on the body whose acceleration is being measured.   Both
issues were far from ideal in the Pioneer missions that were designed for the exploration of
the outer Solar System. Pioneer 10 took 37~years to reach 100~AU and substantial systematic uncertainties on the
measurement of the acceleration vector $\vec{a}(R)$ occurred due to thruster leakages, drag
produced by interplanetary dust and solar wind, and recoils against various forms of radiation
emitted by the spacecraft.

The concept discussed here is based on a low mass ($M=200$~kg) spacecraft propelled by a heavy
rocket.
After a series of planetary flybys designed to gain speed, the spacecraft would then coast, while performing the measurements. For concreteness, we assume that the coast phase takes the spacecraft from $\sim1$--$100$~AU, consistent with a series of flybys via Jupiter, Mars, and finally, Earth. The simple model here conservatively assumes that the position and velocity of the spacecraft is measured once every $\sim 3$~weeks. Likely, Deep Space Network (DSN) ranging will be available for measurement more often than this assumption.  A preliminary and generic flight time calculation is consistent with a total
coast time to 100~AU of seven years~\cite{flight_time}. The sensitivity of the experiment depends somewhat on the details of the coasting trajectory; for most of this paper we consider a polar trajectory, perpendicular to the ecliptic plane of the Solar System. This trajectory reduces the impact of the Kuiper Belt's highly uncertain gravitational pull. Other trajectories are possible, but have different systematics to consider; we discuss the choice of spacecraft trajectories in Sec.~\ref{sec:traj}.

\subsection{Spacecraft}

\begin{figure}[t!]
\includegraphics[width=.45\textwidth]{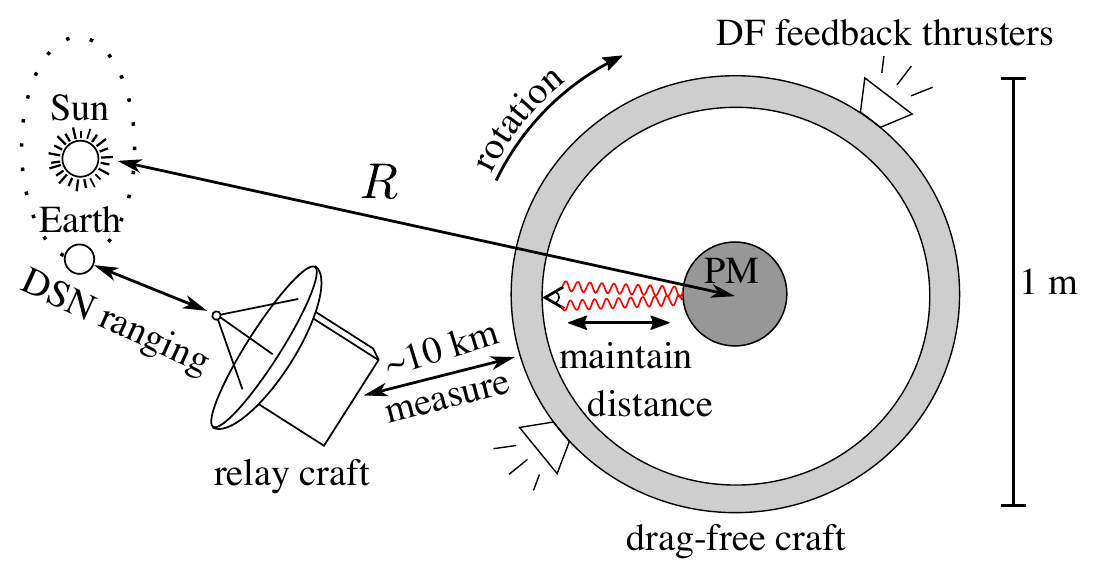}
\caption{Conceptual view of the drag-free (DF) and relay spacecraft. The figure is not to scale, and the DF craft is shown in cross-section. The DF craft uses micro-thrusters to remain centered on the free-falling PM while rotating in a plane perpendicular to the Sun to reduce thermal and gravitational backgrounds. The relay craft carries a high-gain antenna, always pointing towards Earth, for DSN ranging and communication. Communication and ranging between the two spacecraft are performed by small omnidirectional transponders.\label{fig:sketch}}
\end{figure}

\begin{table}[t!]

{
\begin{tabular}{cc}
Parameter   &	Value	    \\
\hline
DF Spacecraft Mass ($M$)	 &  $200$~kg		\\
DF Spacecraft Radius ($r_S$) & $1$~m\\
Experiment Duration ($t$) & $7$~yr\\
PM Mass ($m$) & $10$~kg	    \\
PM Radius 	   &  $5$~cm		     \\
DF Cavity Radius ($r_C$) & $10$~cm \\
Required Thruster Bandwidth	  &   $10^{-2}$~Hz				  \\
Optical Sensing Deadband	($d$)				&		$10~\mu$m					\\
Minimum Correction Period					&  $100$~s				\\
Time Between DSN Measurements	  &  $2 \times 10^{6}$~s     \\
PM Discharge Period								&  $2\times 10^5$~s				\\
Microthruster Fuel Mass	&   $< 50$~g				\\
Angular Velocity	($\omega/2\pi$)    &  $0.1$~Hz   \\
Spacecraft Velocity, Radial ($v$) &	$14$~AU/yr (initial)	  \\
Trailing Spacecraft Distance  &  $10$~km			    \\
RTG Thermal Power Requirement  &  $1$~kW			    \\
\end{tabular}
}
\caption{\label{tab:parameters}Summary of relevant parameters, as outlined in the text. The PM mass assumes the use of platinum for its construction, and the fuel mass required includes only the amount necessary for operating the DF system.  The RTG thermal power refers to the trailing spacecraft and would result in $\sim 50$ W of electrical power.  The drag-free craft may use a substantially smaller RTG or, maybe more likely, receive power transmitted optically from the relay craft. The symbols in parentheses match those used in Table~\ref{tab:errors}.}
\end{table}

Central to the spacecraft design is the use of the drag-free (DF)
technique~\cite{DF}, whereby a feedback system ``flies'' the spacecraft around a ``proof
mass'' (PM) that, to within a very high degree of accuracy, is subject only to gravity.
The PM is stowed and inactive during the initial maneuvering and flyby phases and is only released
and tracked during the coast.
The DF technique was developed in the '60s and initially tested on the U.S. Navy
TRIAD spacecraft~\cite{TRIAD}.
More recently, Gravity Probe B~\cite{GPB} used the DF technique
in a challenging configuration, where each of the quartz rotors of the gyroscopes at the heart
of the experiment were also used as PMs. In 2009, ESA launched the Gravity Field and Steady-State Ocean Circulation
Explorer (GOCE)~\cite{GOCE} that utilized the DF design to map Earth's
gravitational field.  The Laser Interferometer Space Antenna (LISA)~\cite{LISA} also
plans to use the DF technique to establish a highly accurate geodesic network between
PMs located on different (and widely separated) spacecraft. As will become clear, the DF system discussed here is conceptually simpler than those of Gravity Probe B and LISA, although new and unique challenges are presented by the long duration of the flight and the needs of telemetry and ranging over very large distances.

In our model we assume that measurements of the $1/R^2$
law can be carried out over about two orders of magnitude in distance, from $\sim 1$~AU to 100~AU.
While conventional chemical engines would be employed for the flyby phase,
such engines would be jettisoned at the beginning of the coast when micro-thrusters optimized to assist the DF system over the long duration of the flight
would take over.   Standard telemetry using NASA's Deep Space Network (DSN)~\cite{DSN} would
provide range and Doppler data at well known times, from which $\vec{a}(R)$ can be derived.

A conceptual sketch of the spacecraft in the coasting configuration is shown in Fig.~\ref{fig:sketch}.
The ultimate performance of a DF system is limited by the interactions
between the spacecraft (shell) and the PM. These include thermal gradients on the inner surface of the shell, residual gas from spacecraft components, and gravitational forces from asymmetries in the shell. The thermal gradients can be reduced with appropriate insulation, and the residual gas can be reduced by appropriate choices of materials. However, gravity cannot be shielded, so great care has to be
taken to build the spacecraft symmetrically around the cavity hosting the PM. 
In order to minimize the residual gravitational interaction of the spacecraft on the PM
at the maximum offset allowed by the feedback system, it is advantageous to design the
gap between the PM and the inside of the cavity to be larger than found in existing DF implementations. In the conceptual design discussed
here, the cavity and PM are spherical, with the cavity radius $r_C = 10$~cm or more and PM radius $r_{PM} = 5$~cm.  These dimensions are not the result of a careful optimization
but are derived in analogy with previous systems~\cite{large_cavity,large_cavity2} with the constraint of
a modest-size spacecraft.  With such a large gap it is likely that optical ranging would
be appropriate for measuring the position and motion of the PM in its housing.
A (probably different) optical system would also be a candidate for initializing the DF system
after the PM is released at the beginning of the coast. Ultraviolet light-emitting diodes (LEDs) inside the cavity would allow for periodic discharging of the PM, which accumulates charge over time due to interactions with cosmic rays, every few days \cite{UVLED,UVLED2}.

The choice of thrusters to be used for the DF system is constrained by considerations of
reliability, fuel endurance and bandwidth.  In addition, care must be taken to ensure that
fuel consumption does not change the position of the mass center of the
spacecraft, producing an anomalous acceleration of the PM.  The bandwidth $f$ and total
impulse $J$ requirements can be estimated by imposing a deadband of 10~$\mu$m on the centering of the PM with respect to the spacecraft center of mass.
This results in $f \sim 10^{-2}$~Hz and $J= 2 \times 10^{-4}$~Ns.
Electromagnetic thrusters provide high specific impulse, proportional control and,
in the case of field emission electric propulsion thrusters (FEEPs)~\cite{Marcuccio}, bandwidths in excess of 1~Hz.  FEEPs emit and
electromagnetically accelerate Cs, Rb, or In ions, providing thrusts of $10^{-7}$--$10^{-2}$~N with the requisite bandwidth~\cite{Marcuccio}. Using the specifications in Ref.~\cite{Marcuccio}
for a cesium-based FEEP, the calculated fuel consumption is $<50$~g for the duration of the mission.
However, additional fuel is likely needed for attitude control and for adjustments to the
spacecraft rotation. While it is not clear if FEEPs with the required reliability will be available, alternative schemes with nominally similar performance include laser ablative microthrusters~\cite{karg2013investigation} and other types of electromagnetic microthrusters, currently under development~\cite{busek}.

In order to further reduce the effect of interactions with the spacecraft on the PM, rotation of the shell around the PM can be imposed during the coast
phase. Such a rotation spectrally shifts all disturbances produced internally to the
spacecraft in the plane perpendicular to the rotation axis. Since, to first order, only accelerations along the coasting
direction are important, the plane of rotation is chosen to contain the Sun. For the purposes of preliminary calculations, we assume a rotation rate of $\sim 0.1$~Hz, which is sufficient to reduce many nuisance interactions to negligible levels while remaining easily achievable with onboard thrusters.

Due to the rotation of the probe and to the need for a high gain directional antenna for communications from deep space, we envisage the mission to include a second, trailing spacecraft approximately
10~km behind the DF probe, to relay telemetry to and from Earth. The 10~km distance renders the gravitational coupling between the trailing spacecraft and the PM negligible, while being short enough to allow for simple omnidirectional communications. The DF and relay crafts would be docked together during the maneuvering phase of the mission, and then separate and take their relative positions at the beginning of the coast.  The relay spacecraft does not need to accurately maintain its position relative to the DF probe, as long as their relative position and velocity is constantly measured with sufficient precision.  In the rest of this paper we will assume that all such ranging and velocity measurements are the combination of the two segments: Earth--relay craft and relay craft--DF craft.

Radioisotope thermal generator(s) (RTGs) will have to be used to provide power to the spacecraft because of the large heliocentric distance reached.  We expect a conventional RTG rated for 1 kW thermal and 50W electrical power to be sufficient for the relay craft. A smaller RTG or a system to optically transmit power can be used for the drag free probe. Optical transmission of power may benefit from a flight formation in which the two crafts are separated in the plane orthogonal to the Sun--PM axis, so that the center of rotation of the drag-free probe is visible from the relay craft.

The relevant parameters of the proposed spacecraft are summarized in Table~\ref{tab:parameters}.

\subsection{Trajectory}\label{sec:traj}

The spacecraft trajectory is a key factor in determining the sensitivity of the experiment. The trajectory determines the effects of any new physics as well as Solar System backgrounds. For instance, trajectories that remain quite close to the Sun (say, in a bound, nearly circular orbit) have larger effects from new physics, but cannot discriminate between new physics and backgrounds---in this limit, all effects are, to first order, perihelion precessions. Such experiments can set excellent limits with null results (lunar laser ranging (LLR), for example, sets the strongest limit on $|\alpha|$ of any existing experiment) but are unable to directly confirm any signal they may observe. For that reason, we focus on trajectories that traverse a wide range of distance scales, making possible the identification of a signal based on its radial dependence.

The Pioneer and Voyager spacecraft followed such trajectories, close to the ecliptic plane of the Solar System in order to conduct several planetary flybys and observations. However, a trajectory in the ecliptic plane passes through the Kuiper Belt, which has a very poorly constrained mass distribution. The systematic uncertainty introduced by the Kuiper Belt in this case is found to be substantial. A polar trajectory, coasting perpendicular to the ecliptic plane, reduces the effect of the Kuiper Belt while also allowing its mass distribution to be fit with a small number of parameters (see Appendix~\ref{app:kuiper}). We note here that future experimental constraints on the mass distribution of the Kuiper Belt may turn the ecliptic trajectory into a competitive option, but that possibility is not considered further in this work.

Reference~\cite{flight_time} shows that spacecraft velocities of up to 14 AU/yr are achievable with a sequence of planetary flybys. Achieving these high speeds requires careful trajectory designs which we do not attempt here. Instead we present an estimate of sensitivity using a representative trajectory whose final flyby is around Earth, out of the ecliptic, and at an initial speed of 14 AU/yr, reaching 77 AU from the Sun in seven years.

\section{Analysis of systematics}
	\begin{table*}[t!]
		\renewcommand{\arraystretch}{1.3}
	\begin{tabular}{l|cccc|c}
		Source 		& Random? 		& Type	 		& Magnitude		& $\delta R$	& Design Constraint \\
		\hline
		Ranging 	& yes			& pos.			& $1\unit{m}$   & $1\unit{m}$  	& --- \\
		\hline
		Thermal gradient &			&				&				&& \\
		---External & no			& $R^{-2}$ accel.& $a_0 = \frac{4\sigma A T^3\Delta T}{mc}$ & $a_0 t^2 \frac{R_0}{R}$ & $\Delta T\lesssim 3\times 10^{-5} \unit{K}(\frac{300\unit{K}}{T})^3$\\
		---Internal & no			& rot.~accel.	& $a = \frac{4\sigma A T^3\Delta T}{mc}$ & $\frac{a t}{\omega}$ & $\Delta T \lesssim 50\unit{K} (\frac{300\unit{K}}{T})^3 $\\
		\hline
		Charging	& 				&				&				&& \\
		---Electrostatic & $\propto q,d$ & accel.	& $a\simeq \frac{q^2 d}{4\pi\epsilon_0 r_C^3 m}$ & $\frac{1}{2}a t^2$ & $q_\textrm{max}\lesssim 1\times 10^7 e$ \\
		---Lorentz	& $\propto q, B$ & accel.		& $a=\frac{q v B}{m}$ & $\frac{1}{2}a t^2$ & $q_\textrm{max}\lesssim 3\times10^7 e (\frac{1\unit{nT}}{B})$ \\
		\hline
		Self-gravity &				&				&				&& \\
		---Dipole $(x,y)$ & no 		& rot.~accel.	& $a\simeq\frac{G\Delta M}{r_C^2}$	& $\frac{a t}{\omega}$ & $\Delta M \lesssim 0.5\unit{kg}$ \\
		---Dipole $(z)$ & no		& accel.		& $a\simeq\frac{G\Delta M}{r_C^2}$	& $\frac{1}{2} t^2 \delta(a\sin\theta)$ &
			 																									$\delta(\Delta M \sin\theta) \lesssim 6\times 10^{-6}\unit{g}$\\
		---Quadrupole $(Q)$ & $\propto d$ & accel.	& $a\simeq\frac{f G M d}{r_S^3}\ln\frac{r_S}{r_C}$ & $\frac{1}{2} a t^2$ & $f\lesssim 1\times 10^{-4}$\\
		\hline
		Residual gas & yes			& accel.		& $\hat{a}\simeq \frac{\sqrt{pA}}{m}(\mu k T)^{1/4}$	& $\frac{1}{2}\hat{a}t^{3/2}$ & $p<30\unit{Pa}(\frac{10\unit{amu}}{\mu})^{1/2}(\frac{300\unit{K}}{T})^{1/4}$ \\
		\hline
		Solar System uncertainties & no & accel.	& \multicolumn{2}{c|}{see text} 									& ---
	\end{tabular}
	\caption{Statistical and systematic errors constraining the sensitivity of the experiment. ``Random?'' is either `yes' for purely statistical errors, `no' for purely constant systematics, or a list of potentially random quantities that the error depends on. (We conservatively assume that these values are constant and set to their maximum value.) ``Type'' is either `pos.' for an error in the position measurement or `accel.' for a force acting on the PM, optionally with `rot.' to indicate that force is rotating (and hence averaged down over many rotations of the shell). ``Magnitude'' is the parametric size of the effect, and $\delta R$ is the displacement of the PM due to the effect after a time $t$. ``Design Constraint'' gives the required size of various parameters in order to keep $\delta R<1\unit{m}$ after the entire 7 yr coast, assuming the gross parameters given in Table~\ref{tab:parameters}. Each source of error is discussed further in the text.\label{tab:errors}}
	\end{table*}

The fundamental limit to the sensitivity of the experiment is the accuracy with which the position of the PM is measured via DSN ranging. However, there are numerous other sources of error, resulting from external forces which act on the PM. Although the drag-free system eliminates the largest external forces on the PM, forces which act directly on the PM (either external or caused by the shell itself) will still affect the net motion of the spacecraft. These forces must be minimized and well-constrained in order to be able to positively identify a signal.

Some forces (especially those from the shell itself) can be made sufficiently small so as to be negligible compared to the ranging measurement uncertainty. Requiring that the force lead to a displacement of less than $1\unit{m}$ after the full $7\unit{yr}$ coast of the spacecraft translates into a constraint on the design of the shell and the precision of its construction. External forces, such as the gravitational forces from objects in the Solar System, cannot be engineered away. Instead they must be precisely modeled and subtracted from the motion of the PM. We find that the masses and positions of the planets are sufficiently well-known to model and subtract their effect~\cite{CBE,ephemeris}.
The Kuiper Belt, despite its small total mass, has a mostly unknown mass distribution and cannot be subtracted or fit by a single parameter; we find that for a polar trajectory, three fit parameters are required, corresponding to the first three terms in the expansion of the Kuiper Belt's gravitational potential discussed in Appendix~\ref{app:kuiper}.

The sources of error and resulting design constraints are summarized in Table~\ref{tab:errors} and fully detailed below.

The forces acting on the PM that we wish to minimize fall into a few categories based on the resulting displacement of the PM. We find that constant, $R^{-2}$, rotating, and random forces all come into play.
A constant force leads to a displacement $\frac{1}{2}aT^2$ after a coast of time $T$.
A force which falls off with distance (e.g.~as $R^{-2}$) is most important at the beginning of the coast, where it builds up some additional velocity, which then adds to the displacement only linearly; the displacement at the end of the coast is roughly $a_0 R_0 T / v$, where $a_0$ is the acceleration at the starting position $R_0\simeq 1\unit{AU}$ of the coast, and $v\simeq 11 \unit{AU/yr}$ is the average velocity of the spacecraft.
Since the shell will be rotating with angular velocity $\omega\simeq \frac{2\pi}{10\unit{s}}$, the effects of some forces originating from it will average down over many rotations; the components of the force lying in the plane of rotation are much less effective at producing a net displacement.
For $\omega T \gg 1$, the displacement of the PM will be $aT/\omega$.
Finally, the PM may be subject to stochastic forces. If the timescale of the force's fluctuations is much shorter than the coast time, the RMS displacement is $\simeq\hat{a}T^{3/2}$, where $\hat{a}=\sigma_{\bar{a}} \sqrt{\tau}$, $\sigma_{\bar{a}}$ is the standard deviation of the stochastic acceleration, averaged over a time period $\tau$ long enough that successive averages are independent while still shorter than the coast time.
The types of forces, their resulting displacements, and the constraints required to keep their displacements below $1\unit{m}$ are shown in Table~\ref{tab:forcetypes}.
As we will discuss below, these constraints appear achievable with careful design and engineering of the spacecraft; therefore, we will treat the DSN ranging as the limiting source of error for the experiment.

\begin{table}[t!]
	\begin{tabular}{ccc}
		Type & Displacement & Constraint \\ \hline
		constant & $\frac{1}{2}a T^2$ & $a \lesssim 4\times 10^{-17}\unit{m/s^2}$ \\
		$R^{-2}$ & $\simeq\frac{a_0 R_0 T}{v}$ & $a_0\lesssim 2 \times 10^{-15}\unit{m/s^2}$ \\
		rotating & $\frac{aT}{\omega}$ & $a\lesssim 3\times 10^{-9}\unit{m/s^2}$ \\
		stochastic & $\simeq\hat{a}T^{3/2}$ & $\hat{a}\lesssim 3\times 10^{-13}\unit{\frac{m}{s^{2} \sqrt{Hz}}}$
	\end{tabular}
	\caption{Summary of the types of forces acting on the proof mass, along with the resulting displacements and the constraint on their sizes obtained by requiring that the displacement remain below $1\unit{m}$ for the entire coast.\label{tab:forcetypes}}
\end{table}

\subsection{Ranging}

The radial position of the spacecraft needs to be periodically measured with considerable precision during the entire coast phase.	Here we assume that the uncertainty of this ranging measurement is dominated by the performance of the Deep Space Network (DSN)~\cite{DSN}.  The reported range 1$\sigma$ accuracy is 1~m. The DSN can also perform a Doppler measurement of the spacecraft's velocity with an accuracy of 0.1~mm/s; however, successive ranging measurements determine the average velocity more precisely, so Doppler information would contribute little additional information to the fit. For this reason, we only consider ranging measurements in the rest of this paper. Negligible uncertainties of $10^{-16}$ yr/yr are to be expected in the time-base for such measurements, using conventional atomic clocks.
We note here that the trajectory of the coast, discussed in Sec.~\ref{sec:traj}, affects the telemetry and ranging. While a single DSN antenna can be used for the polar coast considered here, a trajectory in the ecliptic plane would require a network of ground based antennas during the later stages of the coast once the round-trip travel time of the DSN signal exceeds a few hours.

The two-spacecraft formation slightly complicates the range measurements.  However, the short distance between the DF and the communication spacecraft should be easily measurable with negligible uncertainty compared to the Earth--spacecraft distance.  An interval of $\sim 3$~weeks between measurements, during the entire coast, is sufficient to achieve the accuracy required.

\subsection{Thermal forces}\label{sec:thermal}

Temperature differences on the inner surface of the DF cavity lead to a net force on the PM. Fairly large $\Delta T$s should be expected from internal components of the DF spacecraft, including electronics and power sources (especially if powered by an RTG). However, \emph{external} sources of temperature differences are more constraining because they do not rotate with the shell; internal sources are discussed at the end of this section.

The flux of solar radiation incident on the shell of the spacecraft leads to a thermal gradient across the outer surface of the shell. This gradient can propagate to the inner surface of the shell, leading to a net force on the PM proportional to $(T+\Delta T)^4-T^4\sim 4T^3\Delta T$, where $T$ is the average temperature of the surface and $\Delta T$ is the temperature differential between the hot and cold sides. Although the shell is rotating, the equilibrium temperature difference and hence direction of the net force does not corotate with the shell but rather maintains its alignment relative to the Sun. This leads to a constraint on the (non-corotating) inner surface temperature gradient of $\Delta T(1\unit{AU})\lesssim 3\times 10^{-5} \unit{K}(\frac{300\unit{K}}{T})^3$, using the fact that the gradient due to the solar flux falls with distance as $R^{-2}$. Assuming the total mass, size, and rotation values from Table~\ref{tab:parameters} and using the heat capacity of mm-thick aluminum as a typical value, the temperature difference on the \emph{outer} surface of the spacecraft at $R=1$~AU will be on the order of 1~K. Therefore the outer and inner surfaces of the shell must be thermally decoupled by some insulating layer in order to prevent a large thermal force on the PM.

\begin{figure}[t!]
	\includegraphics[width=0.45\textwidth]{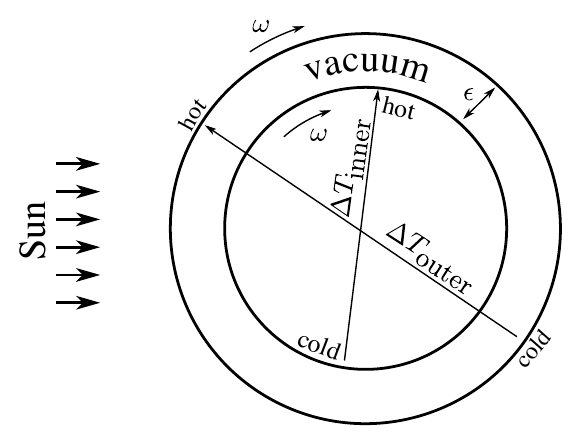}
	\caption{Cross-sectional diagram of the rotating shell with a layer of vacuum insulation. The surfaces facing the vacuum are assumed to have emissivity $\epsilon$. While the material of the shell rotates at $\omega$, the direction of the inner and outer temperature differences $\Delta T$ are fixed relative to the Sun.\label{fig:shell_thermal}}
\end{figure}

Such a small inner $\Delta T$ can in fact be achieved by adding an insulating layer (e.g.~vacuum with low-emissivity coatings) between the outer and inner surfaces of the shell (see Fig.~\ref{fig:shell_thermal}). The temperature difference on the inner surface is then suppressed relative to that on the outer surface if the shell rotates with a period much shorter than the characteristic time of the heat transport between the surfaces. (Note that the inner and outer shells can be mechanically fixed together and corotating; it is the rotation of the shell relative to the Sun that smooths out the temperature.) The heat transport equations are analyzed more fully in Appendix~\ref{app:thermal}. The main result is that in the limit of fast rotations, the suppression from a single layer of vacuum insulation is $\Delta T_\textrm{inner} = \frac{\Delta T_\textrm{outer}}{\omega \tau_\textrm{rad}}$, where $\omega\simeq 0.6/\unitns{s}$ is the angular frequency of the shell's rotation and $\tau_\textrm{rad}\simeq 400\unit{s}\times(\frac{2-\epsilon}{\epsilon})$ for $1\unit{mm}$-thick aluminum at $300\unit{K}$; $\epsilon$ is the emissivity of the surfaces. A large enough suppression may be generated either by making $\epsilon$ small or adding a second insulating layer, as the suppressions from additional layers multiply.

Temperature differences much larger than that created by the Sun will arise from sources internal to the shell, such as electronics and power source; unlike those from the Sun, these differences rotate with the shell and are therefore much less constrained. The temperature differences in the plane of rotation can be as large as $50\unit{K}$ without significantly displacing the PM. The effect of the temperature difference along the axis of rotation is suppressed by choosing the axis of rotation perpendicular to the Sun--spacecraft axis, but in the end will be absorbed by fitting to the transverse acceleration $a_T$ of the PM, which is discussed further in Sec.~\ref{sec:selfgrav}.

\subsection{Charging}
Electromagnetic forces also affect the PM as it charges from cosmic ray collisions. A charged mass inside a spherical (or cylindrical) conducting cavity experiences an acceleration if the mass is offset from the center. Additionally, without magnetic shielding, the PM will experience a Lorentz force from (at minimum) the $\sim$nT magnetic field in the Solar System, as well as any stray fields from the electronics in the shell. If the Solar System magnetic field is dominant, the magnetic and electrostatic effects have a similar strength with the electrostatic effect giving a slightly stronger constraint of $q_\textrm{max}\lesssim 1\times 10^7e$. Assuming a charging rate of 10~protons/s (similar to that expected for the PM in LISA's drag-free system~\cite{LISAA}), discharging the PM using UV LEDs~\cite{UVLED,UVLED2} once every three days is sufficient to keep the resulting electromagnetic forces sufficiently small. Most likely, by discharging more frequently, the stray fields from the electronics will not need to be engineered or shielded all the way down to $\sim$nT, but a detailed analysis of the electronics and associated fields is beyond the scope of this study.

\subsection{Self-gravity}\label{sec:selfgrav}
Gravitational interactions between the shell and PM are not corrected by the drag-free setup and are therefore a potential source of error. To gain an understanding of the typical size and type of gravitational force that can arise, we consider a generic mass distribution $\rho(\vec r)$ for the shell and assume that $\rho(\vec r)=0$ for $|\vec r|<r_C$, the inner radius of the cavity in which the DF system operates. The PM is taken to lie at $\vec r = 0$. Then we can perform an internal multipole expansion of the gravitational potential, so that
\begin{align}
	a_i &= P_i + Q_{ij}r_j + \textrm{higher multipoles},\label{eq:pm_accel_multipoles} \\
	P_i &= G\int d^3 \vec r' \frac{\rho(\vec r') r'_i}{r'^3}, \\
	Q_{ij} &= \frac{1}{2}G\int d^3 \vec r' \frac{\rho (3 r'_i r'_j - r'^2\delta_{ij})}{r'^5},
\end{align}
where we have included only the dipole $P_i$ and quadrupole $Q_{ij}$ because the higher order terms are negligible: the maximum size of the acceleration from $\ell$-th multipole moment is $\frac{GM}{r_C^2}(\frac{d}{r_C})^{\ell-1}$, where d is the PM’s displacement from the center of the cavity. For the parameters assumed in Table~\ref{tab:parameters}, the $\ell=3$ and higher terms are naturally small enough to ignore.

\begin{figure}[t!]
	\includegraphics[width=0.45\textwidth]{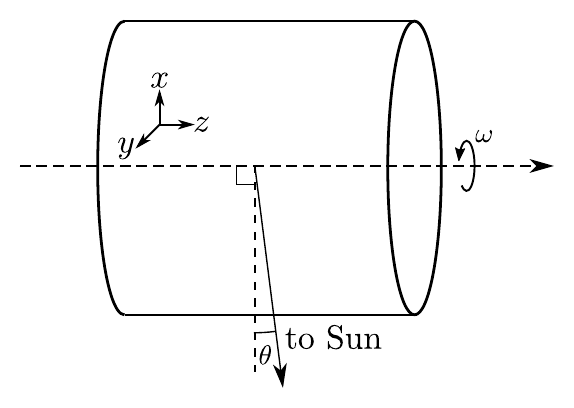}
	\caption{Coordinate system used in Sec.~\ref{sec:selfgrav} for the rotating shell. $\theta$ is the misalignment angle between the plane of rotation and the direction to the Sun.\label{fig:shell_axes}}
\end{figure}

Since the shell is rotating, the multipole moments vary in time. Expressed (as we have) as tensors, they transform like tensors under a time-dependent rotation matrix. However, the PM displacement does not rotate, so the force on the PM from Eq.~(\ref{eq:pm_accel_multipoles}) has a complicated (non-tensorial) transformation. The largest effects come from time-independent forces, so we first consider those. Each multipole moment has $2\ell+1$ independent components and one of those (which can be identified as the $m=0$ part) does not transform under rotations in the $x$--$y$ plane. For the dipole this is simply $P_z$, which leads to a constant force perpendicular to the plane of rotation. Since we have chosen the plane of rotation to contain the Sun, this force does not have a leading-order effect on the PM's motion. However, the plane of rotation must be carefully chosen to ensure that the effect remains small, and the second-order effect from misalignment is still significant. To address this, we leave the constant transverse acceleration $a_T\simeq P_z$ as a free parameter in the fit; then, as long as the spacecraft is pointed accurately, the $a_T$ measured from the fit can be combined with the pointing of the spacecraft to accurately subtract this systematic. Fitting to $a_T$ accounts for all constant, non-rotating forces along the axis of rotation, including the internal temperature difference discussed in Sec.~\ref{sec:thermal}. The coordinate system and the geometry of the situation is shown in Fig.~\ref{fig:shell_axes}.

For the quadrupole, the time-independent component $Q$ appears on the diagonal: $Q_{xx,yy} = Q + \textrm{time-dependent terms}$, $Q_{zz}=-2Q$. The magnitude of the resulting acceleration (in the $x$--$y$ plane) is $a = Q d$ where $d$ is the displacement from the origin. The bound on the size of $Q\lesssim\frac{GM}{r_C^3}$ is much larger than can be ignored, but is also very conservative, so a more careful argument is required. In particular, the bound assumes all of the shell's mass is concentrated near the DF cavity, while a more realistic mass distribution would have most of the mass located towards the outer surface of the craft, changing the parametric dependence of $Q$. For a uniform mass density $\rho$, we have $Q\sim G\rho\ln\frac{r_S}{r_C}\sim\frac{GM}{r_S^3}\ln\frac{r_S}{r_C}$, giving $a\simeq 3\times10^{-13}\unit{m/s^2}$ for the parameters in Table~\ref{tab:parameters}. Precise engineering of the spacecraft shell, on the order of $10^{-4}$ fractional precision on the mass and position of components, is required to reduce this acceleration to a negligible level. The actual form of the spacecraft shell could be based roughly on a zero-$Q$ geometry (such as a uniform-density sphere or cylinder with equal length and diameter) as a starting point and then be shimmed to reduce $Q$ to the necessary level. Furthermore, this estimate of the required precision is still conservative because, with periodic corrections to the offset $d$ provided by the DF system, the displacement will grow more slowly than $t^2$ as assumed here. This argument is only intended to estimate the required precision and establish plausibility; a detailed analysis of the mass distribution and associated uncertainties will be required for any proposed implementation.

The relevant time-dependent components are $P_x$ and $P_y$, but these components are naturally small enough given the stringent constraint on $Q$.

\subsection{Residual gas}
The spacecraft shell will be under a high but finite vacuum. Residual gas will induce a stochastic acceleration of the PM via molecular collisions. We assume the gas to be in thermal equilibrium with the shell and PM, with a temperature $T$, pressure $p$, and typical particle mass $\mu$. Since the size of fluctuations of the average acceleration depends on the averaging timescale, the relevant quantity to calculate is $\hat{a}\equiv \sigma_{\bar{a}}\sqrt{\tau}$ where $\bar{a}$ is the average acceleration over timescale $\tau$ and $\sigma_{\bar{a}}$ denotes the standard deviation of $\bar{a}$. Then, for $\tau$ long enough that successive $\bar{a}$s are statistically independent, the RMS displacement of the PM will be given by $\simeq \hat{a}T^{3/2}$. For the case of gas in the molecular regime, we can take $\tau$ to be the mean time between collisions and $\sigma_{\bar{a}}$ to be typical acceleration imparted by a single collision over time $\tau$. Then it is clear that $\sigma_{\bar{a}}\simeq \frac{pA}{m}$ and $\tau\simeq(nAv)^{-1}$. For an ideal gas $n=\frac{p}{kT}$ and $\left\langle v^2\right\rangle=\frac{kT}{\mu}$, so altogether $\hat{a}\simeq \frac{\sqrt{pA}}{m}(\mu kT)^{1/4}$.
The constraint on the size of a stochastic force from Table~\ref{tab:forcetypes} translates into a limit on the pressure $p\lesssim 10\unit{Pa}$ for 10 amu gas particles at 300 K.

\subsection{Solar System objects}
Objects in our Solar System exert irreducible gravitational forces on the PM. Hence, they must be included in the model of the spacecraft's trajectory. Most objects, including all of the inner planets, are well-measured enough to include in the model with no effect on the experiment's sensitivity~\cite{CBE,ephemeris}. However, the effects of the Sun and the Kuiper Belt are uncertain enough that they must be included as free parameters in the fit. (As a side effect, this experiment will provide the best measurement of the Kuiper Belt's mass distribution; see Sec.~\ref{sec:solarsystemfit}.)
The Sun is always included as a free parameter since for large Yukawa scales $\lambda$ the dominant effect of the new force is to unobservably rescale the mass of the Sun.

Of particular importance is the modeling of the Kuiper Belt. Unlike a point-mass-like planet, the mass distribution of the Kuiper Belt is highly uncertain. In particular, a trajectory which passes directly through the Kuiper Belt in the ecliptic plane is subject to a large systematic uncertainty which is difficult to characterize. One could still set a conservative limit using the data from such a trajectory, but the resulting limit would be much worse than existing limits from planetary motion. However, a polar trajectory does not pass through the Kuiper Belt and allows for a systematic expansion of its gravitational potential (see Appendix~\ref{app:kuiper}). As a result we can include a finite number of parameters in our Solar System model. Under the assumption of a very nearly polar orbit (so that the Kuiper Belt can effectively be averaged azimuthally), we find that the necessary parameters to fit to are (equivalent to) the mass, radius, and offset from the ecliptic plane.

\subsection{Statistical method}\label{sec:statmethod}
To translate the periodic range measurements obtained from the DSN into a measurement of (or limit on) new Yukawa-type forces, we must fit the measurements to a model that includes both the hypothetical new force and all of the known systematic effects. We will consider the simplest case where all of the ranging measurements $X_i$ are independent and Gaussian with equal variances. This case is realized if all of the design constraints in Table~\ref{tab:errors} are satisfied, although random forces can also be accounted for by the fit procedure with a slight generalization. In this case the best-fit (maximum-likelihood) parameters of the model are determined by minimizing
\begin{equation}
	\chi^2(\vec{\theta})=\sum_i \frac{(X_i - \mu(t_i;\vec\theta))^2}{\sigma^2}. \label{eq:chisquared}
\end{equation}
Here $\vec\theta=\{\theta_a\}$ is a vector of the free parameters of the model $\mu(t;\vec\theta)$, which gives the expected range of the spacecraft as a function of time and the model parameters. The standard deviation of each range measurement is $\sigma$. We indicate the best-fit parameters by $\hat\theta$. Then the uncertainty of each of the $\hat\theta$ is encoded in the matrix inverse of the second derivative of $\chi^2$:
\begin{equation}
	(V^{-1})_{ab} = \left.\frac{1}{2}\frac{\partial^2\chi^2(\vec\theta)}{\partial\theta_a\partial\theta_b}\right|_{\vec\theta=\hat\theta},\label{eq:inv_covariance}
\end{equation}
where the inverse on the LHS is a matrix inverse. $V_{ab}$ is the covariance matrix of the best-fit parameters, so that the one-sigma uncertainty in $\hat\theta_a$ is $\sqrt{V_{aa}}$. Of course, this value depends on all of the derivatives of $\chi^2$ via the matrix inverse, and the best-fit parameters may have substantial covariance.

The expected sensitivity of the experiment can be computed approximately by taking the expectation value of Eq.~(\ref{eq:inv_covariance}), leading to
\begin{equation}
		\left\langle(V^{-1})_{ab}\right\rangle = \sum_i\sigma^{-2}\frac{\partial\mu(t_i)}{\partial\theta_a}\frac{\partial\mu(t_i)}{\partial\theta_b},\label{eq:fisher}
\end{equation}
also known as the Fisher information matrix. Note that taking the expectation value of Eq.~(\ref{eq:inv_covariance}) is equivalent to neglecting terms $\propto(X_i-\mu(t_i))\frac{\partial^2\mu}{\partial\theta^2}$, which are suppressed in our case since the $\Delta\theta$ corresponding to $\Delta\mu\sim\sigma$ are well within the linear regime, and so inverting Eq.~(\ref{eq:fisher}) is an excellent approximation to the true expected sensitivity $\left\langle V_{ab}\right\rangle$. Furthermore, computing Eq.~(\ref{eq:fisher}) involves only first derivatives of $\mu$, which is a substantial simplification since $\mu$ must be numerically integrated with very high precision for each value of $\vec\theta$.

It must be noted that the confidence interval obtained from this procedure is only as trustworthy as the model itself, so care must be taken to include all relevant contributions. In particular, a small value of $\chi^2$ at the minimum is not a sufficient condition for the validity of the confidence intervals---one-sigma fluctuations of $\chi^2$ are of size $\sqrt{2N}\simeq 14$ for $N\simeq 100$ measurements, comparable to a \emph{14-sigma} systematic effect. Instead we use $\Delta \chi^2 < 1$ as the criterion for exclusion from the fit.

The free parameters in our simplified Solar System are the masses of the Sun and the Kuiper Belt; the Kuiper Belt $s$ and $z$ moments (defined in Appendix~\ref{app:kuiper}, equivalent to the radius and offset from the ecliptic plane); the initial position and velocity of the spacecraft; and the transverse acceleration of the spacecraft along its rotation axis. The model is then augmented with the parameters of the new physics model under consideration, for instance $\alpha$ for a new Yukawa force with some scale $\lambda$, or $m^{-1}$ for a theory of massive gravity. 

\begin{figure}[t!]
	\includegraphics[width=0.45\textwidth]{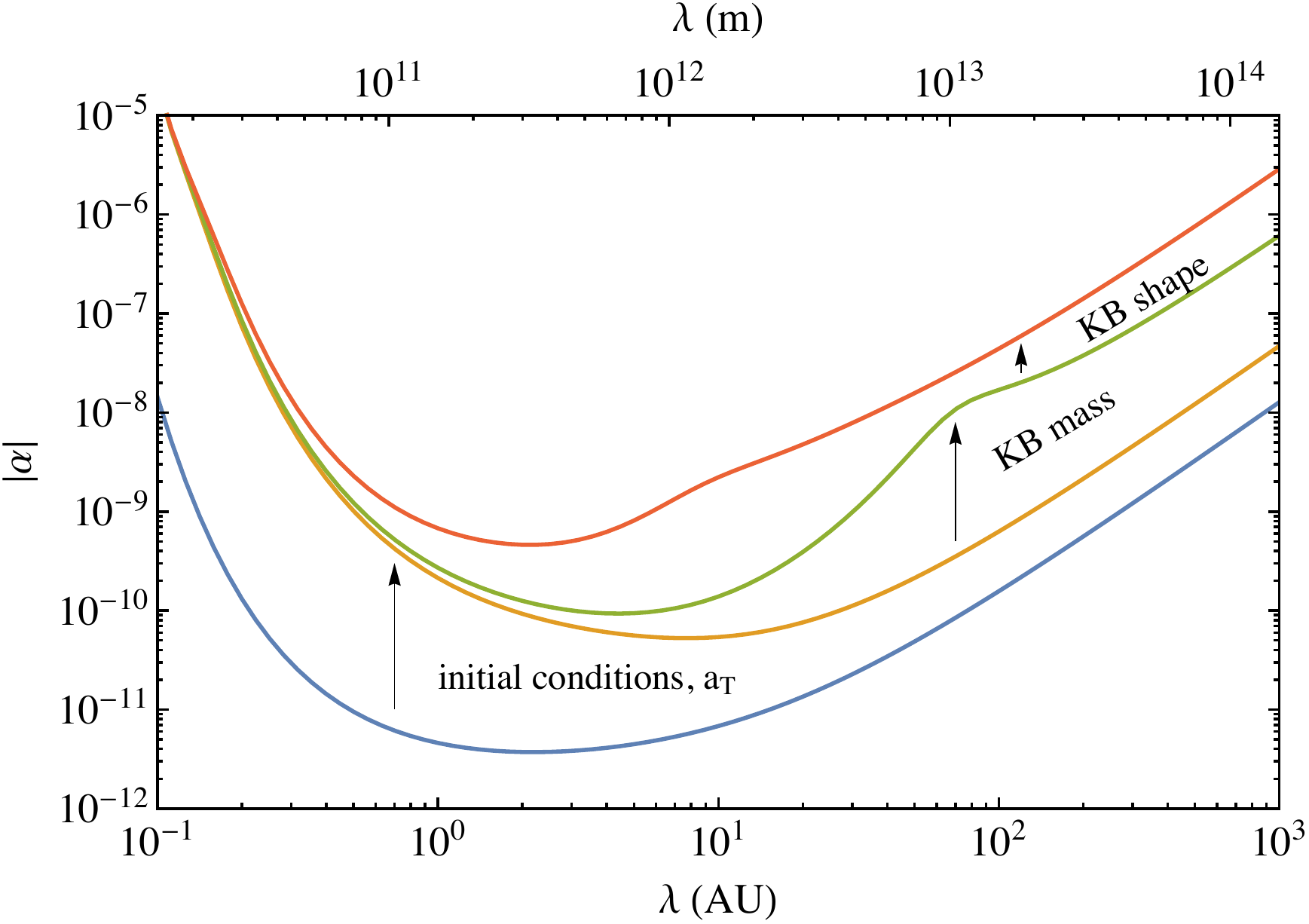}
	\caption{Effect of fit parameters on experiment's sensitivity to a new Yukawa force. The lowest curve results from fixing all parameters except $GM_\textrm{Sun}$ and $\alpha$; from bottom to top we add as free parameters the initial position and velocity and transverse acceleration, mass of the Kuiper Belt, and shape of the Kuiper Belt as described by its radius and offset from the ecliptic plane.\label{fig:parameters}}
\end{figure}

The effects of the various fit parameters on the experiment's sensitivity are shown in Fig.~\ref{fig:parameters} for the case of a new Yukawa force; the results are discussed further in Sec.~\ref{sec:yukawa_results}.

\section{Sensitivity}

\subsection{New forces}\label{sec:yukawa_results}

\begin{figure}[t!]
	\includegraphics[width=0.45\textwidth]{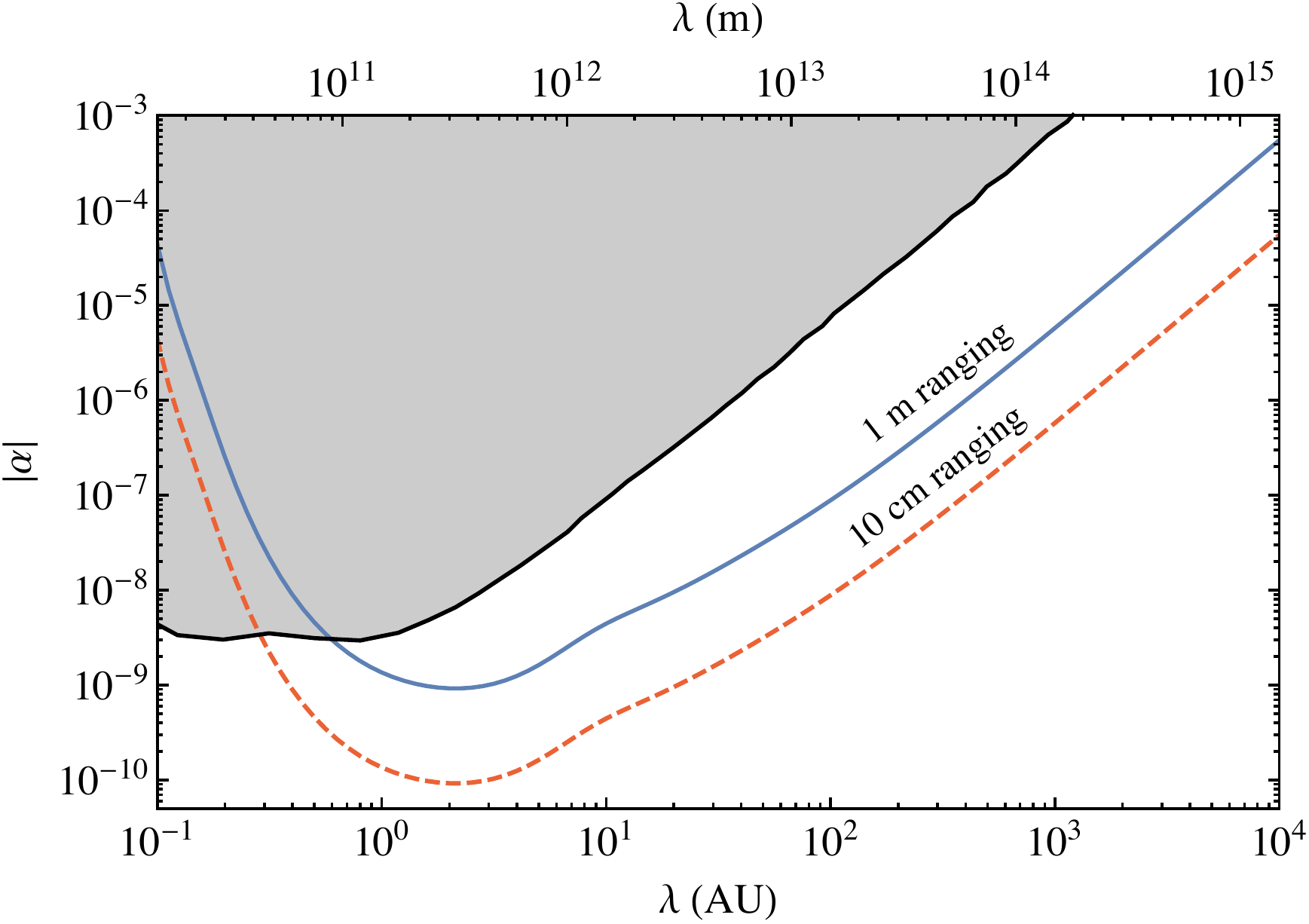}
	\caption{Expected 2$\sigma$ exclusion limit on the strength $\alpha$ of a new Yukawa-type force with range $\lambda$. The blue solid curve is for conservative 1 m ranging precision, while the red dashed curve is an optimistic experiment with 10 cm ranging as the limiting uncertainty. The shaded region is excluded by planetary tests of Kepler's third law.\label{fig:alpha_limit}}
\end{figure}

To compute the sensitivity of the experiment to new Yukawa forces, we compute the expected inverse covariance matrix according to Eq.~(\ref{eq:fisher}), modifying the gravitational potential in accordance with Eq.~(\ref{eq:Yuk}) and including the free parameters of our simplified Solar System as well as the strength $\alpha$ of the Yukawa correction. We present the expected $2\sigma$ limit on $|\alpha|$ as a function of the scale $\lambda$ in Fig.~\ref{fig:alpha_limit}.

At scales $\lambda$ much greater than the distance scale of any particular experiment, the Yukawa correction to Eq.~(\ref{eq:Yuk}) becomes approximately $\delta \Psi\sim-\frac{\alpha GM}{R}(1-\frac{R}{\lambda}+\frac{R^2}{2\lambda^2}+\mathcal{O}(\lambda^{-3}))$. The first term is degenerate with rescaling the source mass, and the second term is an unobservable constant shift in the potential, so that the leading observable correction is a constant radial acceleration $a=\frac{\alpha GM}{2\lambda^2}$. As a result the limit on $|\alpha|$ is generically proportional to $\lambda^2$ at large $\lambda$. At scales $\lambda\gtrsim 100$ AU our expected limit, assuming 1~m ranging, is uniformly two orders of magnitude stronger than the best existing limit (planetary tests of Kepler's third law). Ten-centimeter ranging, which may be possible in the future~\cite{DSN_10cm}, would improve the limit by another order of magnitude if it remained the limiting source of uncertainty.

\subsection{Long-distance modifications of gravity}

\begin{figure}[t!]
	\includegraphics[width=0.45\textwidth]{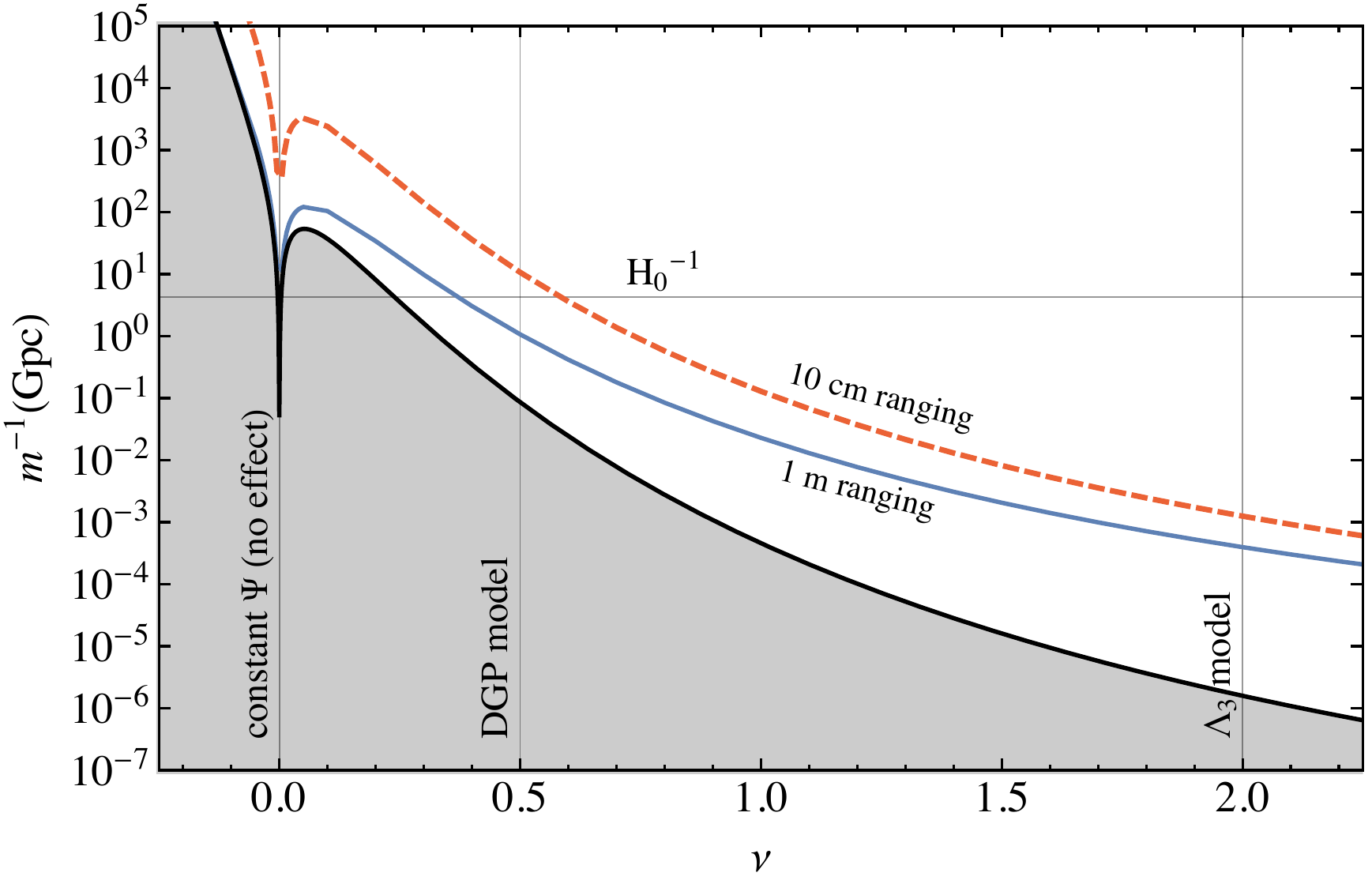}
	\caption{Expected $2\sigma$ lower limit on the graviton Compton wavelength $m^{-1}$ of a new Vainshtein-type power-law contribution to the gravitational potential. For a particular theory with index $\nu$, the experiment will be sensitive to the region below the curve. The vertical lines at $\nu=\frac{1}{2}$ and $\nu=2$ correspond to DGP gravity and ghost-free $\Lambda_3$ massive gravity respectively, and the horizontal line indicates the present-day Hubble scale which is the scale of interest for models that attempt to address the cosmological constant problem. The solid blue curve is for 1 m ranging precision, while the dashed red curve is an optimistic scenario with 10 cm ranging as the limiting uncertainty. The shaded area is excluded by lunar laser ranging. \label{fig:powerlaw_limit}}
\end{figure}

While Yukawa forces are a generic possibility for new physics at long characteristic distance scales, it is generically difficult to construct complete models of Yukawa forces with scales longer than the Earth--Moon distance that our proposed experiment is sensitive to without running afoul of limits on equivalence principle (EP) violation. One class of models (including DGP gravity~\cite{Dvali:2000hr} and some theories of massive gravity~\cite{deRham:2010kj}) that respects EP while also modifying gravity at long distances relies on the Vainshtein mechanism~\cite{Vainshtein:1972sx} to screen the scalar mode of the graviton at short distances, resulting in a nonlinear power law correction to the effective Newtonian potential~\cite{Gruzinov:2001hp,Chkareuli:2011te,Koyama:2011xz,Koyama:2011yg}. To accommodate a variety of models simultaneously, we parametrize the correction to the gravitational potential $\Psi$ as
\begin{equation}
	\delta \Psi \simeq \frac{GM}{R}\left(\frac{R}{R_*}\right)^{\nu+1}\label{eq:powerlaw}
\end{equation}
where $\nu$ is the ``index'' of the power law ($\delta\Psi\propto R^\nu$) and $R_*$ is the Vainshtein radius, the distance below which the scalar graviton is screened. Equation~(\ref{eq:powerlaw}) holds in the screened regime, $R\ll R_*$.  In terms of the graviton mass $m$, $R_*=(GM/m^2)^{1/3}$. (In this section we set $c=\hbar=1$ so that $m^{-1}$ represents a length scale.)

For DGP gravity, $\nu=\frac{1}{2}$ and $m^{-1}\sim\frac{M_\textrm{Pl}^2}{M_5^3}$ is the DGP length scale expressed in terms of the 4D and 5D Planck masses, $M_\textrm{Pl}$ and $M_5$; Equation~(\ref{eq:powerlaw}) reduces to $\delta\Psi\simeq m\sqrt{GMR}$, in agreement with Ref.~\cite{Dvali:2002vf}.  A different power law potential with $\nu=2$ results in the Vainshtein-screened regime of ``ghost-free'' (or ``$\Lambda_3$'') massive gravity~\cite{deRham:2010kj,Chkareuli:2011te}; in this case Eq.~(\ref{eq:powerlaw}) reduces to $\delta\Psi\simeq m^2 R^2$. In both DGP and massive gravity, the value of the graviton Compton wavelength $m^{-1}$ relevant to the cosmological constant problem is today's Hubble scale, $\mathcal{O}(H_0^{-1})\sim 4$~Gpc, corresponding to a graviton mass of $m\sim 10^{-33}$~eV.

The expected sensitivity of our proposed experiment to $m^{-1}$ as a function of $\nu$ is presented in Fig.~\ref{fig:powerlaw_limit}. The experiment is sensitive to length scales below the curves; the index $\nu$ characterizes a particular theory. (Here we compare to lunar laser ranging since it is a direct probe of the potential in the screened regime; Ref.~\cite{Choudhury:2002pu} has considered indirect effects of a graviton mass on weak lensing. They do not directly constrain the parameter space in Fig.~\ref{fig:powerlaw_limit} because they rely on the unscreened regime of massive gravity where the potential cuts off exponentially.) For the particular case of DGP gravity ($\nu=\frac{1}{2}$), our expected $2\sigma$ limit on $m^{-1}$ is 1 Gpc (10 Gpc) for 1 m (10 cm) ranging, compared to the current best limit of $\simeq 100$ Mpc from lunar laser ranging~\cite{Dvali:2002vf,2002nmgm.meet.1797W,Adelberger:2003zx}. Thus, our proposed experiment is sensitive to the most interesting range of DGP parameter space. For steeper power laws (such as the $\nu=2$ potential of $\Lambda_3$ massive gravity; see Fig.~\ref{fig:powerlaw_limit}) our experiment becomes even more sensitive relative to lunar laser ranging due to the longer baseline of our measurements.

\subsection{Modified Newtonian dynamics}

Modified Newtonian dynamics (MOND) invokes deviations from Newton's second law at small accelerations~$a\lesssim a_0 \simeq 10^{-10}\unit{m/s^2}$ as an alternative explanation of galactic rotation curves that does not require the existence of dark matter~\cite{MOND,MONDReview}. However, in this form, MOND has a free functional parameter that interpolates between $F=ma$ for $a\gg a_0$ and $F\propto a^2$ for $a\ll a_0$. The smallest acceleration (due to the Sun) that the spacecraft experiences during its coast is $a\simeq 10^{-6} \unit{m/s^2}$, so we are always in the deeply Newtonian regime of MOND, where predictions are highly dependent on the interpolating function.

However, there are various indirect effects of MOND (due to the external field of the galaxy~\cite{Blanchet:2010it} or the asphericity of the Solar System~\cite{Iorio:2012wv}) that are relatively model-independent. These should manifest themselves in the Newtonian regime as an extra quadrupolar potential $\delta \Psi \simeq Q_2 R^2$. Interpreting the limit on a $\nu=2$ power law from Fig.~\ref{fig:powerlaw_limit} as a quadrupole contribution leads to an expected limit of $Q_2\lesssim 10^{-27}\unit{s^{-2}}$, roughly an order of magnitude smaller than predicted by Ref.~\cite{Blanchet:2010it} for most interpolating functions. (It should be noted that the precise limit on $Q_2$ depends on the orientation of the spacecraft trajectory, e.g.~relative to the Galactic center for the external field effect.) Due to these effects, our experiment would be sensitive to (or able to constrain) broad categories of MOND models.

\subsection{Kuiper Belt measurements}\label{sec:solarsystemfit}

As the precision of our proposed experiment requires fitting to several parameters describing the Kuiper Belt, the experiment also provides a measurement of those parameters. Computing the expected sensitivity (via Eq.~(\ref{eq:fisher})) for 1 m ranging in the absence of any new physics contribution results in a precise measurement of $GM_\textrm{KB}$ with an absolute uncertainty of $\simeq 5\times 10^{-4}GM_\textrm{Earth}$, corresponding to a 0.5\% precision if the measured mass is at its upper bound of roughly $0.1GM_\textrm{Earth}$~\cite{kuiper_mass}. Additionally, the mass-weighted mean radius and offset from the ecliptic plane could each be measured with a precision of about $0.2\unit{AU}\left(\frac{GM_\textrm{KB}}{0.1GM_\textrm{Earth}}\right)$. These would constitute the first direct measurements of the mass distribution of the Kuiper Belt. Combined with optical and infrared observations~\cite{kuiper_dist,cfeps}, knowledge of the mass distribution could constrain the number and distribution of the smallest Kuiper Belt objects.

\section{Conclusions}

We examine the possibility of a space mission to 100~AU dedicated to the precision study of the $1/R^2$ behavior of gravity. Such an experiment would extend the long-distance edge of our knowledge of the gravitational force. 

The $\sim$100~AU baseline enables a much more powerful probe of long-distance modifications of gravity because their effects, relative to ordinary $1/R^2$ gravity, increase with distance. Even assuming a simple new Yukawa-type force, the sensitivity at the longest distance scales would improve by two orders of magnitude over current limits. But in fact, current Yukawa limits are an extrapolation from shorter-distance tests and the experiment described here would be the first direct test of gravity that actually reaches $\sim$100~AU. This is particularly important for validation of various theories of modified or massive gravity that provide alternatives to dark matter or a cosmological constant; in particular, our proposed experiment is sensitive to the interesting parameter space of the DGP model that modifies gravity at the Hubble ($\sim$Gpc) scale and to the effect of the Galactic gravitational field in MOND. A space mission of this type and range is the longest-scale direct test of gravity achievable in the foreseeable future.

\acknowledgments{
We are indebted to the DSN staff at JPL for providing invaluable information on their system and to the
navigation group at SpaceX for discussions. We also thank Savas Dimopoulos, Lorenzo Iorio, Surjeet Rajendran, and Yue Zhao for helpful discussions, and an anonymous referee for several useful comments and for pointing out a numerical error. This work was supported in part by NSF grant PHY-1316706, DOE Early Career Award DE-SC0012012, and the W.M.~Keck Foundation.
}

\appendix

\section{Kuiper Belt parametrization}\label{app:kuiper}

The Kuiper Belt (KB) is a collection of small objects orbiting the Sun beyond Neptune. While optical and infrared surveys constrain the luminosity distribution of the KB, its overall mass is dominated by the least-luminous objects which are not detected by these surveys and is quite poorly constrained~\cite{kuiper_dist,cfeps}. Despite its small mass (less than about $0.1 M_\textrm{Earth}\simeq 3 \times 10^{-7} M_\textrm{Sun}$~\cite{kuiper_mass}) the large uncertainty in its mass and spatial distribution greatly limits our sensitivity to long-range effects.

To parametrize the gravitational effect of the KB on the PM, we perform a systematic expansion of its gravitational potential assuming that the spacecraft does not pass through the KB itself (i.e.~assuming a trajectory somewhat above the ecliptic). We will find later that a very nearly polar trajectory is required for a reasonably simple model of the KB. To perform the expansion, change to cylindrical coordinates $(s,\phi,z)$ with a fixed radial offset $a$:
\begin{equation}
\left\{\begin{aligned}
	x &= (a+s)\cos\phi \\ y&=(a+s)\sin\phi \\ z&=z
\end{aligned}\right..
\end{equation}
In these coordinates the gravitational potential of the KB is given by
\begin{equation}
	\Psi_\textrm{KB}(\vec r) = -G\int \dd \phi' \dd z' \dd s' \frac{(a+s')\rho_\textrm{KB}(s',z',\phi')}{\left|\vec r-\vec r'\right|},
\end{equation}
where $\rho_\textrm{KB}$ is the mass density of the KB. In analogy with a multipole expansion, we now assume $s',z'\ll\left|\vec r-\vec r'\right|$ over the entire support of $\rho_\textrm{KB}$. In principle, one can also expand $\rho_\textrm{KB}$ into a Fourier series in $\phi$ to obtain a complete\footnote{Unlike a multipole or Fourier expansion, this basis is actually \emph{over}complete, and hence not orthogonal. This does not make it any less useful for our purposes here, but it is necessary to ensure that linearly dependent terms are not included as independent fit parameters; such a problem does not arise until third order in the expansion.} series expansion for $\Psi_\textrm{KB}$, with coefficients indexed by the number of powers of $s'$, $z'$, and $e^{i\phi'}$ appearing in the associated term of the integral.

However, we can greatly reduce the number of relevant parameters with two considerations. First, the radar ranging measurement is most sensitive to forces aligned with the Earth-spacecraft direction; for near-polar trajectories, the effects of displacements in the ecliptic plane are suppressed by $\sim \frac{\unit{AU}}{R}$, which is only a few percent or less by the time substantial displacements from the KB are accumulated. This fact allows us to consider only the radial force, and hence the potential only along the trajectory itself. Secondly, if the trajectory is exactly polar, the non-constant Fourier modes of $\rho_\textrm{KB}$ do not contribute to the potential:
\begin{equation}
	\Psi_\textrm{KB}(r \vec{\hat z}) = -G\int\dd z'\dd s' \frac{(a+s')\int\dd\phi'\,\rho_\textrm{KB}(s',z',\phi')}{\sqrt{(a+s')^2+(r-z')^2}},
\end{equation}
so only the azimuthally averaged $\rho_\textrm{KB}$ contributes. The first few terms are
\begin{equation}
	\Psi_\textrm{KB}(r\vec{\hat z}) = -G\left(\frac{M_{00}}{\sqrt{a^2+r^2}}+\frac{r(rM_{10}+aM_{01})}{a(a^2+r^2)^{3/2}}+\ldots\right),
\end{equation}
where $M_{ij}\equiv\int\dd\phi\dd z\dd s\,as^iz^j\rho_\textrm{KB}(s,z,\phi)$. The effects of the $i+j\ge2$ terms are small enough to ignore, so we include only $M_{00}$, $M_{10}$, and $M_{01}$ as fit parameters. (Floating these parameters is equivalent to floating the total mass, mean radius, and z-offset of a thin ring representing the KB.)

\section{Thermal gradients on a rotating shell}\label{app:thermal}

The shell of the spacecraft receives a large flux of energy from the Sun, creating a temperature profile $T_\textrm{out}(\theta,t)$ on the outer surface which is propagated to the inner surface via radiative transfer through a layer of vacuum insulation. The full heat transport equation, including conduction around the inner layer, is
\begin{equation}
	\dot{T}(\theta,t) = \Gamma_\textrm{rad} (T_\textrm{out}(\theta,t)-T(\theta,t)) + \Gamma_\textrm{cond} \frac{\partial^2 T(\theta,t)}{\partial\theta^2},
\end{equation}
where $\Gamma_\textrm{rad}=\frac{4\epsilon\sigma\bar{T}^3}{(2-\epsilon)c\rho d}$ and $\Gamma_\textrm{cond}=\frac{\kappa}{c\rho a^2}$ are the radiative and conductive transfer rates. We have assumed that the temperature fluctuations are small compared to the average temperature $\bar{T}$ so that the radiative term linearizes. Assuming an outer temperature profile of the form $T_\textrm{out}(\theta,t)=\bar{T}+\Delta T_\textrm{out}e^{i(\theta-\omega t)}$ (i.e.~a temperature gradient that rotates around the shell at angular velocity $\omega$), and an inner temperature profile of the form $T(\theta,t)=\bar{T}+a(t) e^{i(\theta-\omega t)}$ (so that $a(t)$ is the complex amplitude of temperature oscillations on the inner surface), we obtain
\begin{equation}
	\dot{a} - i \omega t a = \Gamma_\textrm{rad}(\Delta T_\textrm{out} - a)-\Gamma_\textrm{cond}a.
\end{equation}
In equilibrium, then, $\dot{a}=0$ and we have $a=z\Delta T_\textrm{out}$, with
\begin{equation}
	z\equiv \rho e^{i\phi} =  \frac{\Gamma_\textrm{rad}}{\Gamma_\textrm{rad}+\Gamma_\textrm{cond}-i\omega},
\end{equation}
leading to $T(\theta)=\bar{T}+\rho\Delta T_\textrm{out}\cos(\theta+\phi-\omega t)$ after taking the real part. The suppression factor is then
\begin{equation}
	\frac{\Delta T_\textrm{in}}{\Delta T_\textrm{out}} = \rho = \frac{\Gamma_\textrm{rad}}{\sqrt{(\Gamma_\textrm{rad}+\Gamma_\textrm{cond})^2+\omega^2}},
\end{equation}
which reduces to $\rho\simeq\frac{\Gamma_\textrm{rad}}{|\omega|}$ for fast rotations.

For multiple layers, one has to solve as many coupled transport equations. However, in the limit $\rho\ll 1$, the effect of additional layers is approximately multiplicative; in particular, two layers of vacuum insulation give a $\simeq\rho^2$ suppression.

\bibliography{gravity}

\begin{thebibliography}{58}%
\makeatletter
\providecommand \@ifxundefined [1]{%
 \@ifx{#1\undefined}
}%
\providecommand \@ifnum [1]{%
 \ifnum #1\expandafter \@firstoftwo
 \else \expandafter \@secondoftwo
 \fi
}%
\providecommand \@ifx [1]{%
 \ifx #1\expandafter \@firstoftwo
 \else \expandafter \@secondoftwo
 \fi
}%
\providecommand \natexlab [1]{#1}%
\providecommand \enquote  [1]{``#1''}%
\providecommand \bibnamefont  [1]{#1}%
\providecommand \bibfnamefont [1]{#1}%
\providecommand \citenamefont [1]{#1}%
\providecommand \href@noop [0]{\@secondoftwo}%
\providecommand \href [0]{\begingroup \@sanitize@url \@href}%
\providecommand \@href[1]{\@@startlink{#1}\@@href}%
\providecommand \@@href[1]{\endgroup#1\@@endlink}%
\providecommand \@sanitize@url [0]{\catcode `\\12\catcode `\$12\catcode
  `\&12\catcode `\#12\catcode `\^12\catcode `\_12\catcode `\%12\relax}%
\providecommand \@@startlink[1]{}%
\providecommand \@@endlink[0]{}%
\providecommand \url  [0]{\begingroup\@sanitize@url \@url }%
\providecommand \@url [1]{\endgroup\@href {#1}{\urlprefix }}%
\providecommand \urlprefix  [0]{URL }%
\providecommand \Eprint [0]{\href }%
\providecommand \doibase [0]{http://dx.doi.org/}%
\providecommand \selectlanguage [0]{\@gobble}%
\providecommand \bibinfo  [0]{\@secondoftwo}%
\providecommand \bibfield  [0]{\@secondoftwo}%
\providecommand \translation [1]{[#1]}%
\providecommand \BibitemOpen [0]{}%
\providecommand \bibitemStop [0]{}%
\providecommand \bibitemNoStop [0]{.\EOS\space}%
\providecommand \EOS [0]{\spacefactor3000\relax}%
\providecommand \BibitemShut  [1]{\csname bibitem#1\endcsname}%
\let\auto@bib@innerbib\@empty
\bibitem [{\citenamefont {Adelberger}\ \emph {et~al.}(2003)\citenamefont
  {Adelberger}, \citenamefont {Heckel},\ and\ \citenamefont
  {Nelson}}]{Adelberger:2003zx}%
  \BibitemOpen
  \bibfield  {author} {\bibinfo {author} {\bibfnamefont {E.}~\bibnamefont
  {Adelberger}}, \bibinfo {author} {\bibfnamefont {B.~R.}\ \bibnamefont
  {Heckel}}, \ and\ \bibinfo {author} {\bibfnamefont {A.}~\bibnamefont
  {Nelson}},\ }\href {http://dx.doi.org/10.1146/annurev.nucl.53.041002.110503}
  {\bibfield  {journal} {\bibinfo  {journal} {Ann.Rev.Nucl.Part.Sci.}\ }\textbf
  {\bibinfo {volume} {53}},\ \bibinfo {pages} {77} (\bibinfo {year} {2003})},\
  \Eprint {http://arxiv.org/abs/hep-ph/0307284} {arXiv:hep-ph/0307284 [hep-ph]}
  \BibitemShut {NoStop}%
\bibitem [{\citenamefont {Dimopoulos}\ and\ \citenamefont
  {Giudice}(1996)}]{Dimopoulos:1996kp}%
  \BibitemOpen
  \bibfield  {author} {\bibinfo {author} {\bibfnamefont {S.}~\bibnamefont
  {Dimopoulos}}\ and\ \bibinfo {author} {\bibfnamefont {G.}~\bibnamefont
  {Giudice}},\ }\href {http://dx.doi.org/10.1016/0370-2693(96)00390-5}
  {\bibfield  {journal} {\bibinfo  {journal} {Phys.Lett.}\ }\textbf {\bibinfo
  {volume} {B379}},\ \bibinfo {pages} {105} (\bibinfo {year} {1996})},\ \Eprint
  {http://arxiv.org/abs/hep-ph/9602350} {arXiv:hep-ph/9602350 [hep-ph]}
  \BibitemShut {NoStop}%
\bibitem [{\citenamefont {Arkani-Hamed}\ \emph {et~al.}(1999)\citenamefont
  {Arkani-Hamed}, \citenamefont {Dimopoulos},\ and\ \citenamefont
  {Dvali}}]{ArkaniHamed:1998nn}%
  \BibitemOpen
  \bibfield  {author} {\bibinfo {author} {\bibfnamefont {N.}~\bibnamefont
  {Arkani-Hamed}}, \bibinfo {author} {\bibfnamefont {S.}~\bibnamefont
  {Dimopoulos}}, \ and\ \bibinfo {author} {\bibfnamefont {G.}~\bibnamefont
  {Dvali}},\ }\href {http://dx.doi.org/10.1103/PhysRevD.59.086004} {\bibfield
  {journal} {\bibinfo  {journal} {Phys.Rev.}\ }\textbf {\bibinfo {volume}
  {D59}},\ \bibinfo {pages} {086004} (\bibinfo {year} {1999})},\ \Eprint
  {http://arxiv.org/abs/hep-ph/9807344} {arXiv:hep-ph/9807344 [hep-ph]}
  \BibitemShut {NoStop}%
\bibitem [{\citenamefont {Antoniadis}\ \emph {et~al.}(1998)\citenamefont
  {Antoniadis}, \citenamefont {Arkani-Hamed}, \citenamefont {Dimopoulos},\ and\
  \citenamefont {Dvali}}]{Antoniadis:1998ig}%
  \BibitemOpen
  \bibfield  {author} {\bibinfo {author} {\bibfnamefont {I.}~\bibnamefont
  {Antoniadis}}, \bibinfo {author} {\bibfnamefont {N.}~\bibnamefont
  {Arkani-Hamed}}, \bibinfo {author} {\bibfnamefont {S.}~\bibnamefont
  {Dimopoulos}}, \ and\ \bibinfo {author} {\bibfnamefont {G.}~\bibnamefont
  {Dvali}},\ }\href {http://dx.doi.org/10.1016/S0370-2693(98)00860-0}
  {\bibfield  {journal} {\bibinfo  {journal} {Phys.Lett.}\ }\textbf {\bibinfo
  {volume} {B436}},\ \bibinfo {pages} {257} (\bibinfo {year} {1998})},\ \Eprint
  {http://arxiv.org/abs/hep-ph/9804398} {arXiv:hep-ph/9804398 [hep-ph]}
  \BibitemShut {NoStop}%
\bibitem [{\citenamefont {Arkani-Hamed}\ \emph {et~al.}(1998)\citenamefont
  {Arkani-Hamed}, \citenamefont {Dimopoulos},\ and\ \citenamefont
  {Dvali}}]{ArkaniHamed:1998rs}%
  \BibitemOpen
  \bibfield  {author} {\bibinfo {author} {\bibfnamefont {N.}~\bibnamefont
  {Arkani-Hamed}}, \bibinfo {author} {\bibfnamefont {S.}~\bibnamefont
  {Dimopoulos}}, \ and\ \bibinfo {author} {\bibfnamefont {G.}~\bibnamefont
  {Dvali}},\ }\href {http://dx.doi.org/10.1016/S0370-2693(98)00466-3}
  {\bibfield  {journal} {\bibinfo  {journal} {Phys.Lett.}\ }\textbf {\bibinfo
  {volume} {B429}},\ \bibinfo {pages} {263} (\bibinfo {year} {1998})},\ \Eprint
  {http://arxiv.org/abs/hep-ph/9803315} {arXiv:hep-ph/9803315 [hep-ph]}
  \BibitemShut {NoStop}%
\bibitem [{\citenamefont {Adelberger}\ \emph {et~al.}(2009)\citenamefont
  {Adelberger}, \citenamefont {Battat}, \citenamefont {Currie}, \citenamefont
  {Folkner}, \citenamefont {Gundlach} \emph {et~al.}}]{Decadal_survey}%
  \BibitemOpen
  \bibfield  {author} {\bibinfo {author} {\bibfnamefont {E.~G.}\ \bibnamefont
  {Adelberger}}, \bibinfo {author} {\bibfnamefont {J.}~\bibnamefont {Battat}},
  \bibinfo {author} {\bibfnamefont {D.}~\bibnamefont {Currie}}, \bibinfo
  {author} {\bibfnamefont {W.~M.}\ \bibnamefont {Folkner}}, \bibinfo {author}
  {\bibfnamefont {J.}~\bibnamefont {Gundlach}},  \emph {et~al.},\ }\href@noop
  {} {\  (\bibinfo {year} {2009})},\ \Eprint {http://arxiv.org/abs/0902.3004}
  {arXiv:0902.3004 [gr-qc]} \BibitemShut {NoStop}%
\bibitem [{\citenamefont {Dvali}\ \emph {et~al.}(2000)\citenamefont {Dvali},
  \citenamefont {Gabadadze},\ and\ \citenamefont {Porrati}}]{Dvali:2000hr}%
  \BibitemOpen
  \bibfield  {author} {\bibinfo {author} {\bibfnamefont {G.}~\bibnamefont
  {Dvali}}, \bibinfo {author} {\bibfnamefont {G.}~\bibnamefont {Gabadadze}}, \
  and\ \bibinfo {author} {\bibfnamefont {M.}~\bibnamefont {Porrati}},\ }\href
  {http://dx.doi.org/10.1016/S0370-2693(00)00669-9} {\bibfield  {journal}
  {\bibinfo  {journal} {Phys.Lett.}\ }\textbf {\bibinfo {volume} {B485}},\
  \bibinfo {pages} {208} (\bibinfo {year} {2000})},\ \Eprint
  {http://arxiv.org/abs/hep-th/0005016} {arXiv:hep-th/0005016 [hep-th]}
  \BibitemShut {NoStop}%
\bibitem [{\citenamefont {Dvali}\ \emph
  {et~al.}(2003{\natexlab{a}})\citenamefont {Dvali}, \citenamefont
  {Gabadadze},\ and\ \citenamefont {Shifman}}]{Dvali:2002pe}%
  \BibitemOpen
  \bibfield  {author} {\bibinfo {author} {\bibfnamefont {G.}~\bibnamefont
  {Dvali}}, \bibinfo {author} {\bibfnamefont {G.}~\bibnamefont {Gabadadze}}, \
  and\ \bibinfo {author} {\bibfnamefont {M.}~\bibnamefont {Shifman}},\ }\href
  {http://dx.doi.org/10.1103/PhysRevD.67.044020} {\bibfield  {journal}
  {\bibinfo  {journal} {Phys.Rev.}\ }\textbf {\bibinfo {volume} {D67}},\
  \bibinfo {pages} {044020} (\bibinfo {year} {2003}{\natexlab{a}})},\ \Eprint
  {http://arxiv.org/abs/hep-th/0202174} {arXiv:hep-th/0202174 [hep-th]}
  \BibitemShut {NoStop}%
\bibitem [{\citenamefont {Dvali}\ \emph {et~al.}(2002)\citenamefont {Dvali},
  \citenamefont {Gabadadze},\ and\ \citenamefont {Shifman}}]{Dvali:2002fz}%
  \BibitemOpen
  \bibfield  {author} {\bibinfo {author} {\bibfnamefont {G.}~\bibnamefont
  {Dvali}}, \bibinfo {author} {\bibfnamefont {G.}~\bibnamefont {Gabadadze}}, \
  and\ \bibinfo {author} {\bibfnamefont {M.}~\bibnamefont {Shifman}},\
  }\href@noop {} {\  (\bibinfo {year} {2002})},\ \Eprint
  {http://arxiv.org/abs/hep-th/0208096} {arXiv:hep-th/0208096 [hep-th]}
  \BibitemShut {NoStop}%
\bibitem [{\citenamefont {Arkani-Hamed}\ \emph {et~al.}(2002)\citenamefont
  {Arkani-Hamed}, \citenamefont {Dimopoulos}, \citenamefont {Dvali},\ and\
  \citenamefont {Gabadadze}}]{ArkaniHamed:2002fu}%
  \BibitemOpen
  \bibfield  {author} {\bibinfo {author} {\bibfnamefont {N.}~\bibnamefont
  {Arkani-Hamed}}, \bibinfo {author} {\bibfnamefont {S.}~\bibnamefont
  {Dimopoulos}}, \bibinfo {author} {\bibfnamefont {G.}~\bibnamefont {Dvali}}, \
  and\ \bibinfo {author} {\bibfnamefont {G.}~\bibnamefont {Gabadadze}},\
  }\href@noop {} {\  (\bibinfo {year} {2002})},\ \Eprint
  {http://arxiv.org/abs/hep-th/0209227} {arXiv:hep-th/0209227 [hep-th]}
  \BibitemShut {NoStop}%
\bibitem [{\citenamefont {de~Rham}\ and\ \citenamefont
  {Gabadadze}(2010)}]{deRham:2010ik}%
  \BibitemOpen
  \bibfield  {author} {\bibinfo {author} {\bibfnamefont {C.}~\bibnamefont
  {de~Rham}}\ and\ \bibinfo {author} {\bibfnamefont {G.}~\bibnamefont
  {Gabadadze}},\ }\href {http://dx.doi.org/10.1103/PhysRevD.82.044020}
  {\bibfield  {journal} {\bibinfo  {journal} {Phys.Rev.}\ }\textbf {\bibinfo
  {volume} {D82}},\ \bibinfo {pages} {044020} (\bibinfo {year} {2010})},\
  \Eprint {http://arxiv.org/abs/1007.0443} {arXiv:1007.0443 [hep-th]}
  \BibitemShut {NoStop}%
\bibitem [{\citenamefont {de~Rham}\ \emph
  {et~al.}(2011{\natexlab{a}})\citenamefont {de~Rham}, \citenamefont
  {Gabadadze}, \citenamefont {Heisenberg},\ and\ \citenamefont
  {Pirtskhalava}}]{deRham:2010tw}%
  \BibitemOpen
  \bibfield  {author} {\bibinfo {author} {\bibfnamefont {C.}~\bibnamefont
  {de~Rham}}, \bibinfo {author} {\bibfnamefont {G.}~\bibnamefont {Gabadadze}},
  \bibinfo {author} {\bibfnamefont {L.}~\bibnamefont {Heisenberg}}, \ and\
  \bibinfo {author} {\bibfnamefont {D.}~\bibnamefont {Pirtskhalava}},\ }\href
  {http://dx.doi.org/10.1103/PhysRevD.83.103516} {\bibfield  {journal}
  {\bibinfo  {journal} {Phys.Rev.}\ }\textbf {\bibinfo {volume} {D83}},\
  \bibinfo {pages} {103516} (\bibinfo {year} {2011}{\natexlab{a}})},\ \Eprint
  {http://arxiv.org/abs/1010.1780} {arXiv:1010.1780 [hep-th]} \BibitemShut
  {NoStop}%
\bibitem [{\citenamefont {de~Rham}\ \emph
  {et~al.}(2011{\natexlab{b}})\citenamefont {de~Rham}, \citenamefont
  {Gabadadze},\ and\ \citenamefont {Tolley}}]{deRham:2010kj}%
  \BibitemOpen
  \bibfield  {author} {\bibinfo {author} {\bibfnamefont {C.}~\bibnamefont
  {de~Rham}}, \bibinfo {author} {\bibfnamefont {G.}~\bibnamefont {Gabadadze}},
  \ and\ \bibinfo {author} {\bibfnamefont {A.~J.}\ \bibnamefont {Tolley}},\
  }\href {http://dx.doi.org/10.1103/PhysRevLett.106.231101} {\bibfield
  {journal} {\bibinfo  {journal} {Phys.Rev.Lett.}\ }\textbf {\bibinfo {volume}
  {106}},\ \bibinfo {pages} {231101} (\bibinfo {year} {2011}{\natexlab{b}})},\
  \Eprint {http://arxiv.org/abs/1011.1232} {arXiv:1011.1232 [hep-th]}
  \BibitemShut {NoStop}%
\bibitem [{\citenamefont {D'Amico}\ \emph {et~al.}(2011)\citenamefont
  {D'Amico}, \citenamefont {de~Rham}, \citenamefont {Dubovsky}, \citenamefont
  {Gabadadze}, \citenamefont {Pirtskhalava} \emph {et~al.}}]{D'Amico:2011jj}%
  \BibitemOpen
  \bibfield  {author} {\bibinfo {author} {\bibfnamefont {G.}~\bibnamefont
  {D'Amico}}, \bibinfo {author} {\bibfnamefont {C.}~\bibnamefont {de~Rham}},
  \bibinfo {author} {\bibfnamefont {S.}~\bibnamefont {Dubovsky}}, \bibinfo
  {author} {\bibfnamefont {G.}~\bibnamefont {Gabadadze}}, \bibinfo {author}
  {\bibfnamefont {D.}~\bibnamefont {Pirtskhalava}},  \emph {et~al.},\ }\href
  {http://dx.doi.org/10.1103/PhysRevD.84.124046} {\bibfield  {journal}
  {\bibinfo  {journal} {Phys.Rev.}\ }\textbf {\bibinfo {volume} {D84}},\
  \bibinfo {pages} {124046} (\bibinfo {year} {2011})},\ \Eprint
  {http://arxiv.org/abs/1108.5231} {arXiv:1108.5231 [hep-th]} \BibitemShut
  {NoStop}%
\bibitem [{\citenamefont {de~Rham}(2012)}]{deRham:2012az}%
  \BibitemOpen
  \bibfield  {author} {\bibinfo {author} {\bibfnamefont {C.}~\bibnamefont
  {de~Rham}},\ }\href {http://dx.doi.org/10.1016/j.crhy.2012.04.006} {\bibfield
   {journal} {\bibinfo  {journal} {Comptes Rendus Physique}\ }\textbf {\bibinfo
  {volume} {13}},\ \bibinfo {pages} {666} (\bibinfo {year} {2012})},\ \Eprint
  {http://arxiv.org/abs/1204.5492} {arXiv:1204.5492 [astro-ph.CO]} \BibitemShut
  {NoStop}%
\bibitem [{\citenamefont {Dvali}\ \emph
  {et~al.}(2003{\natexlab{b}})\citenamefont {Dvali}, \citenamefont {Gruzinov},\
  and\ \citenamefont {Zaldarriaga}}]{Dvali:2002vf}%
  \BibitemOpen
  \bibfield  {author} {\bibinfo {author} {\bibfnamefont {G.}~\bibnamefont
  {Dvali}}, \bibinfo {author} {\bibfnamefont {A.}~\bibnamefont {Gruzinov}}, \
  and\ \bibinfo {author} {\bibfnamefont {M.}~\bibnamefont {Zaldarriaga}},\
  }\href {http://dx.doi.org/10.1103/PhysRevD.68.024012} {\bibfield  {journal}
  {\bibinfo  {journal} {Phys.Rev.}\ }\textbf {\bibinfo {volume} {D68}},\
  \bibinfo {pages} {024012} (\bibinfo {year} {2003}{\natexlab{b}})},\ \Eprint
  {http://arxiv.org/abs/hep-ph/0212069} {arXiv:hep-ph/0212069 [hep-ph]}
  \BibitemShut {NoStop}%
\bibitem [{\citenamefont {Berezhiani}\ \emph {et~al.}(2013)\citenamefont
  {Berezhiani}, \citenamefont {Chkareuli},\ and\ \citenamefont
  {Gabadadze}}]{Berezhiani:2013dw}%
  \BibitemOpen
  \bibfield  {author} {\bibinfo {author} {\bibfnamefont {L.}~\bibnamefont
  {Berezhiani}}, \bibinfo {author} {\bibfnamefont {G.}~\bibnamefont
  {Chkareuli}}, \ and\ \bibinfo {author} {\bibfnamefont {G.}~\bibnamefont
  {Gabadadze}},\ }\href {http://dx.doi.org/10.1103/PhysRevD.88.124020}
  {\bibfield  {journal} {\bibinfo  {journal} {Phys.Rev.}\ }\textbf {\bibinfo
  {volume} {D88}},\ \bibinfo {pages} {124020} (\bibinfo {year} {2013})},\
  \Eprint {http://arxiv.org/abs/1302.0549} {arXiv:1302.0549 [hep-th]}
  \BibitemShut {NoStop}%
\bibitem [{\citenamefont {Milgrom}(1983)}]{MOND}%
  \BibitemOpen
  \bibfield  {author} {\bibinfo {author} {\bibfnamefont {M.}~\bibnamefont
  {Milgrom}},\ }\href {http://dx.doi.org/10.1086/161130} {\bibfield  {journal}
  {\bibinfo  {journal} {Astrophys.J.}\ }\textbf {\bibinfo {volume} {270}},\
  \bibinfo {pages} {365} (\bibinfo {year} {1983})}\BibitemShut {NoStop}%
\bibitem [{\citenamefont {Famaey}\ and\ \citenamefont
  {McGaugh}(2012)}]{MONDReview}%
  \BibitemOpen
  \bibfield  {author} {\bibinfo {author} {\bibfnamefont {B.}~\bibnamefont
  {Famaey}}\ and\ \bibinfo {author} {\bibfnamefont {S.}~\bibnamefont
  {McGaugh}},\ }\href {http://dx.doi.org/10.12942/lrr-2012-10} {\bibfield
  {journal} {\bibinfo  {journal} {Living Rev.Rel.}\ }\textbf {\bibinfo {volume}
  {15}},\ \bibinfo {pages} {10} (\bibinfo {year} {2012})},\ \Eprint
  {http://arxiv.org/abs/1112.3960} {arXiv:1112.3960 [astro-ph.CO]} \BibitemShut
  {NoStop}%
\bibitem [{\citenamefont {Blanchet}\ and\ \citenamefont
  {Novak}(2011{\natexlab{a}})}]{Blanchet:2011pv}%
  \BibitemOpen
  \bibfield  {author} {\bibinfo {author} {\bibfnamefont {L.}~\bibnamefont
  {Blanchet}}\ and\ \bibinfo {author} {\bibfnamefont {J.}~\bibnamefont
  {Novak}},\ }\href@noop {} {\  (\bibinfo {year} {2011}{\natexlab{a}})},\
  \Eprint {http://arxiv.org/abs/1105.5815} {arXiv:1105.5815 [astro-ph.CO]}
  \BibitemShut {NoStop}%
\bibitem [{\citenamefont {Agnese}\ \emph {et~al.}(2014)\citenamefont {Agnese}
  \emph {et~al.}}]{CDMS}%
  \BibitemOpen
  \bibfield  {author} {\bibinfo {author} {\bibfnamefont {R.}~\bibnamefont
  {Agnese}} \emph {et~al.} (\bibinfo {collaboration} {SuperCDMS}),\ }\href
  {http://dx.doi.org/10.1103/PhysRevLett.112.241302} {\bibfield  {journal}
  {\bibinfo  {journal} {Phys. Rev. Lett.}\ }\textbf {\bibinfo {volume} {112}},\
  \bibinfo {pages} {241302} (\bibinfo {year} {2014})},\ \Eprint
  {http://arxiv.org/abs/1402.7137} {arXiv:1402.7137 [hep-ex]} \BibitemShut
  {NoStop}%
\bibitem [{\citenamefont {Aprile}\ \emph {et~al.}(2012)\citenamefont {Aprile}
  \emph {et~al.}}]{XENON}%
  \BibitemOpen
  \bibfield  {author} {\bibinfo {author} {\bibfnamefont {E.}~\bibnamefont
  {Aprile}} \emph {et~al.} (\bibinfo {collaboration} {XENON100}),\ }\href
  {http://dx.doi.org/10.1103/PhysRevLett.109.181301} {\bibfield  {journal}
  {\bibinfo  {journal} {Phys. Rev. Lett.}\ }\textbf {\bibinfo {volume} {109}},\
  \bibinfo {pages} {181301} (\bibinfo {year} {2012})},\ \Eprint
  {http://arxiv.org/abs/1207.5988} {arXiv:1207.5988 [astro-ph.CO]} \BibitemShut
  {NoStop}%
\bibitem [{\citenamefont {Akerib}\ \emph {et~al.}(2014)\citenamefont {Akerib}
  \emph {et~al.}}]{LUX}%
  \BibitemOpen
  \bibfield  {author} {\bibinfo {author} {\bibfnamefont {D.}~\bibnamefont
  {Akerib}} \emph {et~al.} (\bibinfo {collaboration} {LUX}),\ }\href
  {http://dx.doi.org/10.1103/PhysRevLett.112.091303} {\bibfield  {journal}
  {\bibinfo  {journal} {Phys.Rev.Lett.}\ }\textbf {\bibinfo {volume} {112}},\
  \bibinfo {pages} {091303} (\bibinfo {year} {2014})},\ \Eprint
  {http://arxiv.org/abs/1310.8214} {arXiv:1310.8214 [astro-ph.CO]} \BibitemShut
  {NoStop}%
\bibitem [{\citenamefont {Hinterbichler}(2012)}]{Hinterbichler:2011tt}%
  \BibitemOpen
  \bibfield  {author} {\bibinfo {author} {\bibfnamefont {K.}~\bibnamefont
  {Hinterbichler}},\ }\href {http://dx.doi.org/10.1103/RevModPhys.84.671}
  {\bibfield  {journal} {\bibinfo  {journal} {Rev. Mod. Phys.}\ }\textbf
  {\bibinfo {volume} {84}},\ \bibinfo {pages} {671} (\bibinfo {year} {2012})},\
  \Eprint {http://arxiv.org/abs/1105.3735} {arXiv:1105.3735 [hep-th]}
  \BibitemShut {NoStop}%
\bibitem [{\citenamefont {Talmadge}\ \emph {et~al.}(1988)\citenamefont
  {Talmadge}, \citenamefont {Berthias}, \citenamefont {Hellings},\ and\
  \citenamefont {Standish}}]{talmadge}%
  \BibitemOpen
  \bibfield  {author} {\bibinfo {author} {\bibfnamefont {C.}~\bibnamefont
  {Talmadge}}, \bibinfo {author} {\bibfnamefont {J.~P.}\ \bibnamefont
  {Berthias}}, \bibinfo {author} {\bibfnamefont {R.~W.}\ \bibnamefont
  {Hellings}}, \ and\ \bibinfo {author} {\bibfnamefont {E.~M.}\ \bibnamefont
  {Standish}},\ }\href {http://dx.doi.org/10.1103/PhysRevLett.61.1159}
  {\bibfield  {journal} {\bibinfo  {journal} {Phys. Rev. Lett.}\ }\textbf
  {\bibinfo {volume} {61}},\ \bibinfo {pages} {1159} (\bibinfo {year}
  {1988})}\BibitemShut {NoStop}%
\bibitem [{\citenamefont {Turyshev}\ and\ \citenamefont
  {Toth}(2010)}]{PioneerAnomaly}%
  \BibitemOpen
  \bibfield  {author} {\bibinfo {author} {\bibfnamefont {S.~G.}\ \bibnamefont
  {Turyshev}}\ and\ \bibinfo {author} {\bibfnamefont {V.~T.}\ \bibnamefont
  {Toth}},\ }\href {http://dx.doi.org/10.12942/lrr-2010-4} {\bibfield
  {journal} {\bibinfo  {journal} {Living Rev.Rel.}\ }\textbf {\bibinfo {volume}
  {13}},\ \bibinfo {pages} {4} (\bibinfo {year} {2010})},\ \Eprint
  {http://arxiv.org/abs/1001.3686} {arXiv:1001.3686 [gr-qc]} \BibitemShut
  {NoStop}%
\bibitem [{\citenamefont {Anderson}\ \emph {et~al.}(2002)\citenamefont
  {Anderson}, \citenamefont {Laing}, \citenamefont {Lau}, \citenamefont {Liu},
  \citenamefont {Nieto} \emph {et~al.}}]{Anderson:2001sg}%
  \BibitemOpen
  \bibfield  {author} {\bibinfo {author} {\bibfnamefont {J.~D.}\ \bibnamefont
  {Anderson}}, \bibinfo {author} {\bibfnamefont {P.~A.}\ \bibnamefont {Laing}},
  \bibinfo {author} {\bibfnamefont {E.~L.}\ \bibnamefont {Lau}}, \bibinfo
  {author} {\bibfnamefont {A.~S.}\ \bibnamefont {Liu}}, \bibinfo {author}
  {\bibfnamefont {M.~M.}\ \bibnamefont {Nieto}},  \emph {et~al.},\ }\href
  {http://dx.doi.org/10.1103/PhysRevD.65.082004} {\bibfield  {journal}
  {\bibinfo  {journal} {Phys.Rev.}\ }\textbf {\bibinfo {volume} {D65}},\
  \bibinfo {pages} {082004} (\bibinfo {year} {2002})},\ \Eprint
  {http://arxiv.org/abs/gr-qc/0104064} {arXiv:gr-qc/0104064 [gr-qc]}
  \BibitemShut {NoStop}%
\bibitem [{\citenamefont {Turyshev}\ \emph {et~al.}(2011)\citenamefont
  {Turyshev}, \citenamefont {Toth}, \citenamefont {Ellis},\ and\ \citenamefont
  {Markwardt}}]{Pioneer_resolved}%
  \BibitemOpen
  \bibfield  {author} {\bibinfo {author} {\bibfnamefont {S.~G.}\ \bibnamefont
  {Turyshev}}, \bibinfo {author} {\bibfnamefont {V.~T.}\ \bibnamefont {Toth}},
  \bibinfo {author} {\bibfnamefont {J.}~\bibnamefont {Ellis}}, \ and\ \bibinfo
  {author} {\bibfnamefont {C.~B.}\ \bibnamefont {Markwardt}},\ }\href
  {http://dx.doi.org/10.1103/PhysRevLett.107.081103} {\bibfield  {journal}
  {\bibinfo  {journal} {Phys.Rev.Lett.}\ }\textbf {\bibinfo {volume} {107}},\
  \bibinfo {pages} {081103} (\bibinfo {year} {2011})},\ \Eprint
  {http://arxiv.org/abs/1107.2886} {arXiv:1107.2886 [gr-qc]} \BibitemShut
  {NoStop}%
\bibitem [{\citenamefont {{Mewaldt}}\ \emph {et~al.}(1995)\citenamefont
  {{Mewaldt}}, \citenamefont {{Kangas}}, \citenamefont {{Kerridge}},\ and\
  \citenamefont {{Neugebauer}}}]{flight_time}%
  \BibitemOpen
  \bibfield  {author} {\bibinfo {author} {\bibfnamefont {R.~A.}\ \bibnamefont
  {{Mewaldt}}}, \bibinfo {author} {\bibfnamefont {J.}~\bibnamefont {{Kangas}}},
  \bibinfo {author} {\bibfnamefont {S.~J.}\ \bibnamefont {{Kerridge}}}, \ and\
  \bibinfo {author} {\bibfnamefont {M.}~\bibnamefont {{Neugebauer}}},\
  }\href@noop {} {\bibfield  {journal} {\bibinfo  {journal} {Acta
  Astronautica}\ }\textbf {\bibinfo {volume} {35}},\ \bibinfo {pages} {267}
  (\bibinfo {year} {1995})}\BibitemShut {NoStop}%
\bibitem [{\citenamefont {DeBra}(1997)}]{DF}%
  \BibitemOpen
  \bibfield  {author} {\bibinfo {author} {\bibfnamefont {D.}~\bibnamefont
  {DeBra}},\ }\href {http://dx.doi.org/10.1088/0264-9381/14/6/026} {\bibfield
  {journal} {\bibinfo  {journal} {Class.Quant.Grav.}\ }\textbf {\bibinfo
  {volume} {14}},\ \bibinfo {pages} {1549} (\bibinfo {year}
  {1997})}\BibitemShut {NoStop}%
\bibitem [{\citenamefont {{Staff of the Space Department}}\ and\ \citenamefont
  {{Staff of the Guidance and Control Laboratory}}(1974)}]{TRIAD}%
  \BibitemOpen
  \bibfield  {author} {\bibinfo {author} {\bibnamefont {{Staff of the Space
  Department}}}\ and\ \bibinfo {author} {\bibnamefont {{Staff of the Guidance
  and Control Laboratory}}},\ }\href {http://dx.doi.org/10.2514/3.62146}
  {\bibfield  {journal} {\bibinfo  {journal} {Journal of Spacecraft and
  Rockets}\ }\textbf {\bibinfo {volume} {11}},\ \bibinfo {pages} {637}
  (\bibinfo {year} {1974})}\BibitemShut {NoStop}%
\bibitem [{\citenamefont {Everitt}\ \emph {et~al.}(2011)\citenamefont
  {Everitt}, \citenamefont {DeBra}, \citenamefont {Parkinson}, \citenamefont
  {Turneaure}, \citenamefont {Conklin} \emph {et~al.}}]{GPB}%
  \BibitemOpen
  \bibfield  {author} {\bibinfo {author} {\bibfnamefont {C.}~\bibnamefont
  {Everitt}}, \bibinfo {author} {\bibfnamefont {D.}~\bibnamefont {DeBra}},
  \bibinfo {author} {\bibfnamefont {B.}~\bibnamefont {Parkinson}}, \bibinfo
  {author} {\bibfnamefont {J.}~\bibnamefont {Turneaure}}, \bibinfo {author}
  {\bibfnamefont {J.}~\bibnamefont {Conklin}},  \emph {et~al.},\ }\href
  {http://dx.doi.org/10.1103/PhysRevLett.106.221101} {\bibfield  {journal}
  {\bibinfo  {journal} {Phys.Rev.Lett.}\ }\textbf {\bibinfo {volume} {106}},\
  \bibinfo {pages} {221101} (\bibinfo {year} {2011})},\ \Eprint
  {http://arxiv.org/abs/1105.3456} {arXiv:1105.3456 [gr-qc]} \BibitemShut
  {NoStop}%
\bibitem [{\citenamefont {{Drinkwater}}\ \emph {et~al.}(2003)\citenamefont
  {{Drinkwater}}, \citenamefont {{Floberghagen}}, \citenamefont {{Haagmans}},
  \citenamefont {{Muzi}},\ and\ \citenamefont {{Popescu}}}]{GOCE}%
  \BibitemOpen
  \bibfield  {author} {\bibinfo {author} {\bibfnamefont {M.~R.}\ \bibnamefont
  {{Drinkwater}}}, \bibinfo {author} {\bibfnamefont {R.}~\bibnamefont
  {{Floberghagen}}}, \bibinfo {author} {\bibfnamefont {R.}~\bibnamefont
  {{Haagmans}}}, \bibinfo {author} {\bibfnamefont {D.}~\bibnamefont {{Muzi}}},
  \ and\ \bibinfo {author} {\bibfnamefont {A.}~\bibnamefont {{Popescu}}},\
  }\href {http://dx.doi.org/10.1023/A:1026104216284} {\bibfield  {journal}
  {\bibinfo  {journal} {Space Science Reviews}\ }\textbf {\bibinfo {volume}
  {108}},\ \bibinfo {pages} {419} (\bibinfo {year} {2003})}\BibitemShut
  {NoStop}%
\bibitem [{\citenamefont {Sallusti}(2009)}]{LISA}%
  \BibitemOpen
  \bibfield  {author} {\bibinfo {author} {\bibfnamefont {M.}~\bibnamefont
  {Sallusti}},\ }\href
  {http://lisa.nasa.gov/Documentation/LISA-MSE-DD-0001_v1.1.pdf} {\emph
  {\bibinfo {title} {Payload Preliminary Design Description}}},\ \bibinfo
  {type} {Tech. Rep.}\ \bibinfo {number} {LISA-MSE-DD-0001}\ (\bibinfo
  {institution} {LISA Project},\ \bibinfo {year} {2009})\BibitemShut {NoStop}%
\bibitem [{\citenamefont {Pham}\ \emph {et~al.}(2015)\citenamefont {Pham} \emph
  {et~al.}}]{DSN}%
  \BibitemOpen
  \bibfield  {author} {\bibinfo {author} {\bibfnamefont {T.}~\bibnamefont
  {Pham}} \emph {et~al.},\ }\href
  {http://deepspace.jpl.nasa.gov/dsndocs/810-005/} {\emph {\bibinfo {title}
  {DSN Telecommunications Link Design Handbook}}},\ \bibinfo {type} {Tech.
  Rep.}\ \bibinfo {number} {810-005}\ (\bibinfo  {institution} {Jet Propulsion
  Laboratory},\ \bibinfo {year} {2015})\BibitemShut {NoStop}%
\bibitem [{\citenamefont {Israel}\ \emph {et~al.}(1973)\citenamefont {Israel}
  \emph {et~al.}}]{large_cavity}%
  \BibitemOpen
  \bibfield  {author} {\bibinfo {author} {\bibfnamefont {G.}~\bibnamefont
  {Israel}} \emph {et~al.},\ }\href@noop {} {\bibfield  {journal} {\bibinfo
  {journal} {ESRO MS(74) 4}\ }\textbf {\bibinfo {volume} {1}},\ \bibinfo
  {pages} {2} (\bibinfo {year} {1973})}\BibitemShut {NoStop}%
\bibitem [{\citenamefont {Roth}(1975)}]{large_cavity2}%
  \BibitemOpen
  \bibfield  {author} {\bibinfo {author} {\bibfnamefont {E.}~\bibnamefont
  {Roth}},\ }\href
  {http://dx.doi.org/http://dx.doi.org/10.1016/0094-5765(75)90068-5} {\bibfield
   {journal} {\bibinfo  {journal} {Acta Astronautica}\ }\textbf {\bibinfo
  {volume} {2}},\ \bibinfo {pages} {543} (\bibinfo {year} {1975})}\BibitemShut
  {NoStop}%
\bibitem [{\citenamefont {{Sun}}\ \emph {et~al.}(2006)\citenamefont {{Sun}},
  \citenamefont {{Allard}}, \citenamefont {{Buchman}}, \citenamefont
  {{Williams}},\ and\ \citenamefont {{Byer}}}]{UVLED}%
  \BibitemOpen
  \bibfield  {author} {\bibinfo {author} {\bibfnamefont {K.-X.}\ \bibnamefont
  {{Sun}}}, \bibinfo {author} {\bibfnamefont {B.}~\bibnamefont {{Allard}}},
  \bibinfo {author} {\bibfnamefont {S.}~\bibnamefont {{Buchman}}}, \bibinfo
  {author} {\bibfnamefont {S.}~\bibnamefont {{Williams}}}, \ and\ \bibinfo
  {author} {\bibfnamefont {R.~L.}\ \bibnamefont {{Byer}}},\ }\href
  {http://dx.doi.org/10.1088/0264-9381/23/8/S19} {\bibfield  {journal}
  {\bibinfo  {journal} {Classical and Quantum Gravity}\ }\textbf {\bibinfo
  {volume} {23}},\ \bibinfo {pages} {141} (\bibinfo {year} {2006})}\BibitemShut
  {NoStop}%
\bibitem [{\citenamefont {{Balakrishnan}}\ \emph {et~al.}(2012)\citenamefont
  {{Balakrishnan}}, \citenamefont {{Sun}} \emph {et~al.}}]{UVLED2}%
  \BibitemOpen
  \bibfield  {author} {\bibinfo {author} {\bibfnamefont {K.}~\bibnamefont
  {{Balakrishnan}}}, \bibinfo {author} {\bibfnamefont {K.-X.}\ \bibnamefont
  {{Sun}}},  \emph {et~al.},\ }in\ \href@noop {} {\emph {\bibinfo {booktitle}
  {39th COSPAR Scientific Assembly}}},\ \bibinfo {series} {COSPAR Meeting},
  Vol.~\bibinfo {volume} {39}\ (\bibinfo {year} {2012})\ p.~\bibinfo {pages}
  {90},\ \Eprint {http://arxiv.org/abs/1202.0585} {arXiv:1202.0585
  [physics.ins-det]} \BibitemShut {NoStop}%
\bibitem [{\citenamefont {Marcuccio}\ \emph {et~al.}(1998)\citenamefont
  {Marcuccio}, \citenamefont {Genovese},\ and\ \citenamefont
  {Andrenucci}}]{Marcuccio}%
  \BibitemOpen
  \bibfield  {author} {\bibinfo {author} {\bibfnamefont {S.}~\bibnamefont
  {Marcuccio}}, \bibinfo {author} {\bibfnamefont {A.}~\bibnamefont {Genovese}},
  \ and\ \bibinfo {author} {\bibfnamefont {M.}~\bibnamefont {Andrenucci}},\
  }\href {http://dx.doi.org/10.2514/2.5340} {\bibfield  {journal} {\bibinfo
  {journal} {Journal of Propulsion and Power}\ }\textbf {\bibinfo {volume}
  {14}},\ \bibinfo {pages} {774} (\bibinfo {year} {1998})}\BibitemShut
  {NoStop}%
\bibitem [{\citenamefont {Karg}\ and\ \citenamefont
  {Fedotov}(2013)}]{karg2013investigation}%
  \BibitemOpen
  \bibfield  {author} {\bibinfo {author} {\bibfnamefont {S.}~\bibnamefont
  {Karg}}\ and\ \bibinfo {author} {\bibfnamefont {V.}~\bibnamefont {Fedotov}},\
  }in\ \href
  {http://odas2013.onera.fr/ODAS2013/sites/sites.onera.fr.ODAS2013/files/u4/KARG_Investigation%20of%20laser-ablative%20micropropulsion%20as%20an%20alternative%20thruster%20concept%20for%20precise%20satellite%20attitude%20and%20orbit%20control.pdf}
  {\emph {\bibinfo {booktitle} {ONERA-DLR Aerospace Symposium}}}\ (\bibinfo
  {address} {Palaiseu, France},\ \bibinfo {year} {2013})\ pp.\ \bibinfo {pages}
  {27--29}\BibitemShut {NoStop}%
\bibitem [{bus()}]{busek}%
  \BibitemOpen
  \href {http://www.busek.com} {}\bibinfo {howpublished} {See e.g.
  \url{http://www.busek.com}.}\BibitemShut {Stop}%
\bibitem [{\citenamefont {Luzum}\ \emph {et~al.}()\citenamefont {Luzum} \emph
  {et~al.}}]{CBE}%
  \BibitemOpen
  \bibfield  {author} {\bibinfo {author} {\bibfnamefont {B.}~\bibnamefont
  {Luzum}} \emph {et~al.},\ }\href
  {http://maia.usno.navy.mil/NSFA/NSFA_cbe.html} {\emph {\bibinfo {title}
  {Astronomical Constants: Current Best Estimates}}},\ \bibinfo {type} {Tech.
  Rep.}\ (\bibinfo  {institution} {{NSFA Working Group}})\BibitemShut {NoStop}%
\bibitem [{\citenamefont {{Folkner}}(2011)}]{ephemeris}%
  \BibitemOpen
  \bibfield  {author} {\bibinfo {author} {\bibfnamefont {W.~M.}\ \bibnamefont
  {{Folkner}}},\ }in\ \href
  {http://syrte.obspm.fr/jsr/journees2010/pdf/Folkner.pdf} {\emph {\bibinfo
  {booktitle} {Journ{\'e}es Syst{\`e}mes de R{\'e}f{\'e}rence Spatio-temporels
  2010}}},\ \bibinfo {editor} {edited by\ \bibinfo {editor} {\bibfnamefont
  {N.}~\bibnamefont {{Capitaine}}}}\ (\bibinfo {year} {2011})\ pp.\ \bibinfo
  {pages} {43--48}\BibitemShut {NoStop}%
\bibitem [{\citenamefont {Bender}\ \emph {et~al.}(1998)\citenamefont {Bender}
  \emph {et~al.}}]{LISAA}%
  \BibitemOpen
  \bibfield  {author} {\bibinfo {author} {\bibfnamefont {P.}~\bibnamefont
  {Bender}} \emph {et~al.},\ }\href
  {http://lisa.nasa.gov/Documentation/ppa2.08.pdf} {\emph {\bibinfo {title}
  {LISA Pre-Phase A Report}}},\ \bibinfo {type} {Tech. Rep.}\ (\bibinfo
  {institution} {LISA Project},\ \bibinfo {year} {1998})\BibitemShut {NoStop}%
\bibitem [{\citenamefont {{L. Teitelbaum, JPL}}()}]{DSN_10cm}%
  \BibitemOpen
  \bibfield  {author} {\bibinfo {author} {\bibnamefont {{L. Teitelbaum,
  JPL}}},\ }\href@noop {} {}\bibinfo {howpublished} {private
  communication}\BibitemShut {NoStop}%
\bibitem [{\citenamefont {Vainshtein}(1972)}]{Vainshtein:1972sx}%
  \BibitemOpen
  \bibfield  {author} {\bibinfo {author} {\bibfnamefont {A.~I.}\ \bibnamefont
  {Vainshtein}},\ }\href {http://dx.doi.org/10.1016/0370-2693(72)90147-5}
  {\bibfield  {journal} {\bibinfo  {journal} {Phys. Lett.}\ }\textbf {\bibinfo
  {volume} {B39}},\ \bibinfo {pages} {393} (\bibinfo {year}
  {1972})}\BibitemShut {NoStop}%
\bibitem [{\citenamefont {Gruzinov}(2005)}]{Gruzinov:2001hp}%
  \BibitemOpen
  \bibfield  {author} {\bibinfo {author} {\bibfnamefont {A.}~\bibnamefont
  {Gruzinov}},\ }\href {http://dx.doi.org/10.1016/j.newast.2004.12.001}
  {\bibfield  {journal} {\bibinfo  {journal} {New Astron.}\ }\textbf {\bibinfo
  {volume} {10}},\ \bibinfo {pages} {311} (\bibinfo {year} {2005})},\ \Eprint
  {http://arxiv.org/abs/astro-ph/0112246} {arXiv:astro-ph/0112246 [astro-ph]}
  \BibitemShut {NoStop}%
\bibitem [{\citenamefont {Chkareuli}\ and\ \citenamefont
  {Pirtskhalava}(2012)}]{Chkareuli:2011te}%
  \BibitemOpen
  \bibfield  {author} {\bibinfo {author} {\bibfnamefont {G.}~\bibnamefont
  {Chkareuli}}\ and\ \bibinfo {author} {\bibfnamefont {D.}~\bibnamefont
  {Pirtskhalava}},\ }\href {http://dx.doi.org/10.1016/j.physletb.2012.05.030}
  {\bibfield  {journal} {\bibinfo  {journal} {Phys. Lett.}\ }\textbf {\bibinfo
  {volume} {B713}},\ \bibinfo {pages} {99} (\bibinfo {year} {2012})},\ \Eprint
  {http://arxiv.org/abs/1105.1783} {arXiv:1105.1783 [hep-th]} \BibitemShut
  {NoStop}%
\bibitem [{\citenamefont {Koyama}\ \emph
  {et~al.}(2011{\natexlab{a}})\citenamefont {Koyama}, \citenamefont {Niz},\
  and\ \citenamefont {Tasinato}}]{Koyama:2011xz}%
  \BibitemOpen
  \bibfield  {author} {\bibinfo {author} {\bibfnamefont {K.}~\bibnamefont
  {Koyama}}, \bibinfo {author} {\bibfnamefont {G.}~\bibnamefont {Niz}}, \ and\
  \bibinfo {author} {\bibfnamefont {G.}~\bibnamefont {Tasinato}},\ }\href
  {http://dx.doi.org/10.1103/PhysRevLett.107.131101} {\bibfield  {journal}
  {\bibinfo  {journal} {Phys. Rev. Lett.}\ }\textbf {\bibinfo {volume} {107}},\
  \bibinfo {pages} {131101} (\bibinfo {year} {2011}{\natexlab{a}})},\ \Eprint
  {http://arxiv.org/abs/1103.4708} {arXiv:1103.4708 [hep-th]} \BibitemShut
  {NoStop}%
\bibitem [{\citenamefont {Koyama}\ \emph
  {et~al.}(2011{\natexlab{b}})\citenamefont {Koyama}, \citenamefont {Niz},\
  and\ \citenamefont {Tasinato}}]{Koyama:2011yg}%
  \BibitemOpen
  \bibfield  {author} {\bibinfo {author} {\bibfnamefont {K.}~\bibnamefont
  {Koyama}}, \bibinfo {author} {\bibfnamefont {G.}~\bibnamefont {Niz}}, \ and\
  \bibinfo {author} {\bibfnamefont {G.}~\bibnamefont {Tasinato}},\ }\href
  {http://dx.doi.org/10.1103/PhysRevD.84.064033} {\bibfield  {journal}
  {\bibinfo  {journal} {Phys. Rev.}\ }\textbf {\bibinfo {volume} {D84}},\
  \bibinfo {pages} {064033} (\bibinfo {year} {2011}{\natexlab{b}})},\ \Eprint
  {http://arxiv.org/abs/1104.2143} {arXiv:1104.2143 [hep-th]} \BibitemShut
  {NoStop}%
\bibitem [{\citenamefont {Choudhury}\ \emph {et~al.}(2004)\citenamefont
  {Choudhury}, \citenamefont {Joshi}, \citenamefont {Mahajan},\ and\
  \citenamefont {McKellar}}]{Choudhury:2002pu}%
  \BibitemOpen
  \bibfield  {author} {\bibinfo {author} {\bibfnamefont {S.~R.}\ \bibnamefont
  {Choudhury}}, \bibinfo {author} {\bibfnamefont {G.~C.}\ \bibnamefont
  {Joshi}}, \bibinfo {author} {\bibfnamefont {S.}~\bibnamefont {Mahajan}}, \
  and\ \bibinfo {author} {\bibfnamefont {B.~H.~J.}\ \bibnamefont {McKellar}},\
  }\href {http://dx.doi.org/10.1016/j.astropartphys.2004.04.001} {\bibfield
  {journal} {\bibinfo  {journal} {Astropart. Phys.}\ }\textbf {\bibinfo
  {volume} {21}},\ \bibinfo {pages} {559} (\bibinfo {year} {2004})},\ \Eprint
  {http://arxiv.org/abs/hep-ph/0204161} {arXiv:hep-ph/0204161 [hep-ph]}
  \BibitemShut {NoStop}%
\bibitem [{\citenamefont {{Williams}}\ \emph {et~al.}(2002)\citenamefont
  {{Williams}}, \citenamefont {{Boggs}}, \citenamefont {{Dickey}},\ and\
  \citenamefont {{Folkner}}}]{2002nmgm.meet.1797W}%
  \BibitemOpen
  \bibfield  {author} {\bibinfo {author} {\bibfnamefont {J.~G.}\ \bibnamefont
  {{Williams}}}, \bibinfo {author} {\bibfnamefont {D.~H.}\ \bibnamefont
  {{Boggs}}}, \bibinfo {author} {\bibfnamefont {J.~O.}\ \bibnamefont
  {{Dickey}}}, \ and\ \bibinfo {author} {\bibfnamefont {W.~M.}\ \bibnamefont
  {{Folkner}}},\ }in\ \href {http://dx.doi.org/10.1142/9789812777386_0389}
  {\emph {\bibinfo {booktitle} {The Ninth Marcel Grossmann Meeting}}},\
  \bibinfo {editor} {edited by\ \bibinfo {editor} {\bibfnamefont {V.~G.}\
  \bibnamefont {{Gurzadyan}}}, \bibinfo {editor} {\bibfnamefont {R.~T.}\
  \bibnamefont {{Jantzen}}}, \ and\ \bibinfo {editor} {\bibfnamefont
  {R.}~\bibnamefont {{Ruffini}}}}\ (\bibinfo {year} {2002})\ pp.\ \bibinfo
  {pages} {1797--1798}\BibitemShut {NoStop}%
\bibitem [{\citenamefont {Blanchet}\ and\ \citenamefont
  {Novak}(2011{\natexlab{b}})}]{Blanchet:2010it}%
  \BibitemOpen
  \bibfield  {author} {\bibinfo {author} {\bibfnamefont {L.}~\bibnamefont
  {Blanchet}}\ and\ \bibinfo {author} {\bibfnamefont {J.}~\bibnamefont
  {Novak}},\ }\href {http://dx.doi.org/10.1111/j.1365-2966.2010.18076.x}
  {\bibfield  {journal} {\bibinfo  {journal} {Mon. Not. Roy. Astron. Soc.}\
  }\textbf {\bibinfo {volume} {412}},\ \bibinfo {pages} {2530} (\bibinfo {year}
  {2011}{\natexlab{b}})},\ \Eprint {http://arxiv.org/abs/1010.1349}
  {arXiv:1010.1349 [astro-ph.CO]} \BibitemShut {NoStop}%
\bibitem [{\citenamefont {Iorio}(2013)}]{Iorio:2012wv}%
  \BibitemOpen
  \bibfield  {author} {\bibinfo {author} {\bibfnamefont {L.}~\bibnamefont
  {Iorio}},\ }\href {http://dx.doi.org/10.1088/0264-9381/30/16/165018}
  {\bibfield  {journal} {\bibinfo  {journal} {Class. Quant. Gravit.}\ }\textbf
  {\bibinfo {volume} {30}},\ \bibinfo {pages} {165018} (\bibinfo {year}
  {2013})},\ \Eprint {http://arxiv.org/abs/1211.3688} {arXiv:1211.3688 [gr-qc]}
  \BibitemShut {NoStop}%
\bibitem [{\citenamefont {Iorio}(2007)}]{kuiper_mass}%
  \BibitemOpen
  \bibfield  {author} {\bibinfo {author} {\bibfnamefont {L.}~\bibnamefont
  {Iorio}},\ }\href {http://dx.doi.org/10.1111/j.1365-2966.2006.11384.x}
  {\bibfield  {journal} {\bibinfo  {journal} {Mon.Not.Roy.Astron.Soc.}\
  }\textbf {\bibinfo {volume} {375}},\ \bibinfo {pages} {1311} (\bibinfo {year}
  {2007})},\ \Eprint {http://arxiv.org/abs/gr-qc/0609023} {arXiv:gr-qc/0609023
  [gr-qc]} \BibitemShut {NoStop}%
\bibitem [{\citenamefont {{Kavelaars}}\ \emph {et~al.}(2008)\citenamefont
  {{Kavelaars}}, \citenamefont {{Jones}}, \citenamefont {{Gladman}},
  \citenamefont {{Parker}},\ and\ \citenamefont {{Petit}}}]{kuiper_dist}%
  \BibitemOpen
  \bibfield  {author} {\bibinfo {author} {\bibfnamefont {J.}~\bibnamefont
  {{Kavelaars}}}, \bibinfo {author} {\bibfnamefont {L.}~\bibnamefont
  {{Jones}}}, \bibinfo {author} {\bibfnamefont {B.}~\bibnamefont {{Gladman}}},
  \bibinfo {author} {\bibfnamefont {J.~W.}\ \bibnamefont {{Parker}}}, \ and\
  \bibinfo {author} {\bibfnamefont {J.-M.}\ \bibnamefont {{Petit}}},\ }\enquote
  {\bibinfo {title} {{The Orbital and Spatial Distribution of the Kuiper
  Belt}},}\ in\ \href {http://www.cfeps.net/Publications_files/7032.pdf} {\emph
  {\bibinfo {booktitle} {The Solar System Beyond Neptune}}},\ \bibinfo {editor}
  {edited by\ \bibinfo {editor} {\bibfnamefont {M.~A.}\ \bibnamefont
  {{Barucci}}}, \bibinfo {editor} {\bibfnamefont {H.}~\bibnamefont
  {{Boehnhardt}}}, \bibinfo {editor} {\bibfnamefont {D.~P.}\ \bibnamefont
  {{Cruikshank}}}, \bibinfo {editor} {\bibfnamefont {A.}~\bibnamefont
  {{Morbidelli}}}, \ and\ \bibinfo {editor} {\bibfnamefont {R.}~\bibnamefont
  {{Dotson}}}}\ (\bibinfo {year} {2008})\ pp.\ \bibinfo {pages}
  {59--69}\BibitemShut {NoStop}%
\bibitem [{\citenamefont {{Petit}}\ \emph {et~al.}(2011)\citenamefont
  {{Petit}}, \citenamefont {{Kavelaars}}, \citenamefont {{Gladman}},
  \citenamefont {{Jones}}, \citenamefont {{Parker}}, \citenamefont {{Van
  Laerhoven}}, \citenamefont {{Nicholson}}, \citenamefont {{Mars}},
  \citenamefont {{Rousselot}}, \citenamefont {{Mousis}}, \citenamefont
  {{Marsden}}, \citenamefont {{Bieryla}}, \citenamefont {{Taylor}},
  \citenamefont {{Ashby}}, \citenamefont {{Benavidez}}, \citenamefont {{Campo
  Bagatin}},\ and\ \citenamefont {{Bernabeu}}}]{cfeps}%
  \BibitemOpen
  \bibfield  {author} {\bibinfo {author} {\bibfnamefont {J.-M.}\ \bibnamefont
  {{Petit}}}, \bibinfo {author} {\bibfnamefont {J.~J.}\ \bibnamefont
  {{Kavelaars}}}, \bibinfo {author} {\bibfnamefont {B.~J.}\ \bibnamefont
  {{Gladman}}}, \bibinfo {author} {\bibfnamefont {R.~L.}\ \bibnamefont
  {{Jones}}}, \bibinfo {author} {\bibfnamefont {J.~W.}\ \bibnamefont
  {{Parker}}}, \bibinfo {author} {\bibfnamefont {C.}~\bibnamefont {{Van
  Laerhoven}}}, \bibinfo {author} {\bibfnamefont {P.}~\bibnamefont
  {{Nicholson}}}, \bibinfo {author} {\bibfnamefont {G.}~\bibnamefont {{Mars}}},
  \bibinfo {author} {\bibfnamefont {P.}~\bibnamefont {{Rousselot}}},  \emph
  {et~al.},\ }\href {http://dx.doi.org/10.1088/0004-6256/142/4/131} {\bibfield
  {journal} {\bibinfo  {journal} {Astron.J.}\ }\textbf {\bibinfo {volume}
  {142}},\ \bibinfo {eid} {131} (\bibinfo {year} {2011})},\ \Eprint
  {http://arxiv.org/abs/1108.4836} {arXiv:1108.4836 [astro-ph.EP]} \BibitemShut
  {NoStop}%
\end{thebibliography}%

\end{document}